\documentclass[prd,aps,preprint,nofootinbib,amsmath,amssymb,eqsecnum]{revtex4-2}

\usepackage[T1]{fontenc}
\usepackage[utf8]{inputenc}
\usepackage{booktabs}
\usepackage{array}[=2016-10-06]
\usepackage{xcolor}
\usepackage{tikz}
\usetikzlibrary{arrows.meta,positioning,shapes.geometric}
\usepackage[colorlinks=true,allcolors=blue]{hyperref}

\newcommand{\W}{\mathcal{W}}

\newcommand{\tomega}{\tilde{\omega}}
\newcommand{\p}{\tilde{\phi}}
\newcommand{\G}{\tilde{G}}

\newcommand{\ip}[1]{\int_{#1}}

\makeatletter
\def\@fnsymbol#1{\ensuremath{\ifcase#1\or \dagger\or \ddagger\or \mathsection\or
  \mathparagraph\or \|\or \dagger\dagger\or \ddagger\ddagger\else\@ctrerr\fi}}
\makeatother

\begin{document}

\title{Reconstructing the Goldberg--Palatini Action from Noether Identities}

\author{F. T. Brandt}
\email{fbrandt@usp.br}
\author{Augusto C. B. Carvalho}
\email{augusto.burlacchini@usp.br}
\affiliation{Instituto de F\'{\i}sica, Universidade de S\~ao Paulo, S\~ao Paulo, SP 05508-090, Brazil}

\date{\today}

\begin{abstract}
We investigate the reconstruction of the first-order gravitational action from the Noether
identities associated with diffeomorphism invariance. The independent fields are a symmetric
perturbation $\phi^{\mu\nu}$ of the contravariant metric density and a trace-adjusted
connection $G^{\lambda}{}_{\mu\nu}$, whose transformation laws are prescribed. For a local
action polynomial in these fields, with the gravitational coupling as its only dimensionful
constant and each additional field carrying one power of it, a dilatation, which is a
particular diffeomorphism, shows that in any dimension $d>2$ the action terminates at most at
cubic order, its only possible cubic term being a derivative-free $\phi GG$ vertex. Expanding the kernels of the resulting Lorentz-covariant ansatz in tensor
bases gives 31 coefficients for $d\geq4$. The Noether identities with general
diffeomorphisms, expanded order by order in the fields, yield a linear system that eliminates
the $\phi\phi$ kernel and determines the remaining coefficients up to an overall
normalization, recovering the Goldberg--Palatini action. Since the connection can be
eliminated exactly, this also fixes the complete second-order action, with its infinite set
of graviton vertices. The construction requires no choice of energy--momentum tensor, the
source of ambiguity in the self-coupling approach. The kernel bases, the identities and the
solution are given in appendices, and the symbolic computations are publicly available.
\end{abstract}

\maketitle

\section{Introduction}
\label{sec:introduction}

The extent to which gravitational dynamics can be recovered from the consistency of an
interacting massless spin-two field is a longstanding question. In the self-coupling
construction, which goes back to Feynman~\cite{Feynman1995} and was completed by
Deser~\cite{Deser1970}, the use of first-order variables~\cite{Palatini1919,Ferraris1982} allows the nonlinear completion to
be expressed through a cubic interaction, avoiding the infinite expansion of the metric
formulation. Discussions of this approach have also emphasized the
assumptions entering the construction, the ambiguities in the energy--momentum tensor, and the
treatment of boundary terms~\cite{Padmanabhan2008,Deser2010}. These issues motivate a
formulation in which the constraints imposed by gauge symmetry can be examined directly.

Local gauge invariance implies differential identities among the Euler--Lagrange derivatives
of the action. These are identities of Noether's second theorem and hold off shell. Once the
field transformations and a class of admissible actions have been specified, they can be used
as algebraic conditions on unknown interaction coefficients. This reverses the usual direction
of the calculation: instead of deriving identities from a known action, one uses the
identities to determine which actions are compatible with the prescribed transformations.
This strategy was used in Ref.~\cite{BrandtElias2006} to determine the scalar--graviton
vertex in noncommutative space, where the closed form of the action is not available, from
the invariance under a restricted class of coordinate transformations.

In this work we apply this strategy to the first-order gravitational variables $\phi^{\mu\nu}$
and $G^{\lambda}{}_{\mu\nu}$, the perturbation of the contravariant metric density and a
trace-adjusted connection treated as an independent field. The
same strategy applies in the metric formulation, where the field-dependent part of the
transformation is essential: invariance under the linearized transformations alone fixes
only part of each vertex, whereas the full transformation determines the three-graviton
vertex from the quadratic term, up to total derivatives~\cite{BrandtFrenkel1996}. The
program does not terminate there, however, since new vertices appear at every order. The
first-order choice is adapted to the polynomial structure of the Goldberg--Palatini
action~\cite{Goldberg1958,Palatini1919,Ferraris1982} and to its use in perturbative
gravity~\cite{BrandtMcKeon2016}. We first show, using a coordinate dilatation as a particular
diffeomorphism, that a local action polynomial in these variables, with each additional field
carrying one power of the coupling and no other dimensionful constant, terminates at most at
cubic order, its only possible cubic term being a derivative-free interaction $\phi GG$, linear
in $\phi$ and quadratic in $G$ as in the Goldberg--Palatini action. We then expand the kernels
of this class in tensor bases and impose the Noether identities with general diffeomorphisms
order by order in the fields; the determination of the action reduces to a finite system of
linear constraints on 31 coefficients (for $d\geq4$), which is solved completely. Because the field equation of $G$ is algebraic, the
connection can then be eliminated exactly, so the finite determination of the first-order
action also fixes the infinite set of graviton vertices of the metric
formulation~\cite{BrandtMcKeon2016}.

Our objective is to determine the freedom remaining in this class after the Noether
identities are imposed. Locality, Lorentz covariance and parity, the transformation laws, the
field content, the coupling-power assignment that bounds the number of derivatives, and
polynomiality are inputs to the construction; the restriction of the cubic interaction is
derived from them, and the uniqueness statement concerns this specified class of actions and does not assert that
diffeomorphism invariance alone selects Einstein gravity among all local gravitational
theories.

The construction complements the self-coupling approach of
Refs.~\cite{Deser1970,Deser2010}. Self-coupling combines the conserved current of global
translations with the consistency of the gauge identities of the spin-two field, and thus
requires the choice of an energy--momentum tensor, the point at issue in
Ref.~\cite{Padmanabhan2008}. Its recent extension to metric-affine theories~\cite{Delhom2023}
identifies this choice as the source of the ambiguities, notes that in the first-order
construction of Ref.~\cite{Deser1970} the connection is an auxiliary field that is not itself
bootstrapped, and observes that the existing arguments establish existence rather than
uniqueness. In the present work no energy--momentum tensor is constructed: the off-shell identities of
Noether's second theorem are imposed directly, the geometrical input being retained in the
transformation laws; the variations of both independent fields enter the identities; and
uniqueness is established within the stated class. The termination at cubic order, a
property of the solution in the first-order construction of Ref.~\cite{Deser1970}, appears here
as a necessary condition on any polynomial action in these variables. The idea of using the consistency of the
gauge identities instead of self-coupling goes back to Wald~\cite{Wald1986}, who analyzed the
consistent nonlinear extensions through identities among the field equations; here the
nonlinear transformations are prescribed and the compatible action is determined within the
polynomial class. The result concerns the bulk action up to total derivatives and does not
determine boundary terms.

\begin{figure}[htb]
\centering
\begin{tikzpicture}[
  node distance=5mm and 6mm,
  box/.style={draw, rounded corners, align=center, minimum height=9mm, inner sep=5pt, font=\small},
  res/.style={box, fill=black!5},
  arr/.style={-{Latex[length=2mm]}, thick}]
\node[box] (fields) {fields $\phi$, $G$\\Eqs.~\eqref{eq:dphi}--\eqref{eq:dG}};
\node[box, right=of fields] (closure) {allowed sectors, Sec.~\ref{sec:closure}\\$N\le3$, $\phi GG$};
\node[box, right=of closure] (ansatz) {ansatz \eqref{eq:ansatz}\\31 coefficients};
\node[box, right=of ansatz] (ident) {identities \eqref{eq:noether}\\Table~\ref{tab:identities}};
\node[res, below=of ident] (khh) {$\W^{31}$: $K^{\phi\phi}=0$};
\node[res, left=of khh] (s2) {$\W^{\phi},\W^{G}$: $S_2$ up to\\two parameters};
\node[res, left=of s2] (s3) {$\W^{32}$: one parameter,\\$K^{\phi GG}$ fixed};
\node[box, below=of s3] (result) {action \eqref{eq:result}};
\node[box, right=of result] (check) {$\W^{33},\W^{34}$: no further constraints;\\reduction to metric form};
\draw[arr] (fields) -- (closure);
\draw[arr] (closure) -- (ansatz);
\draw[arr] (ansatz) -- (ident);
\draw[arr] (ident) -- (khh);
\draw[arr] (khh) -- (s2);
\draw[arr] (s2) -- (s3);
\draw[arr] (s3) -- (result);
\draw[arr] (result) -- (check);
\end{tikzpicture}
\caption{The determination of the action. Inputs and the allowed sectors of
Section~\ref{sec:closure} on the top row; the sequential solution of
Table~\ref{tab:flow} on the middle row; output and checks on the bottom row.}
\label{fig:flow}
\end{figure}
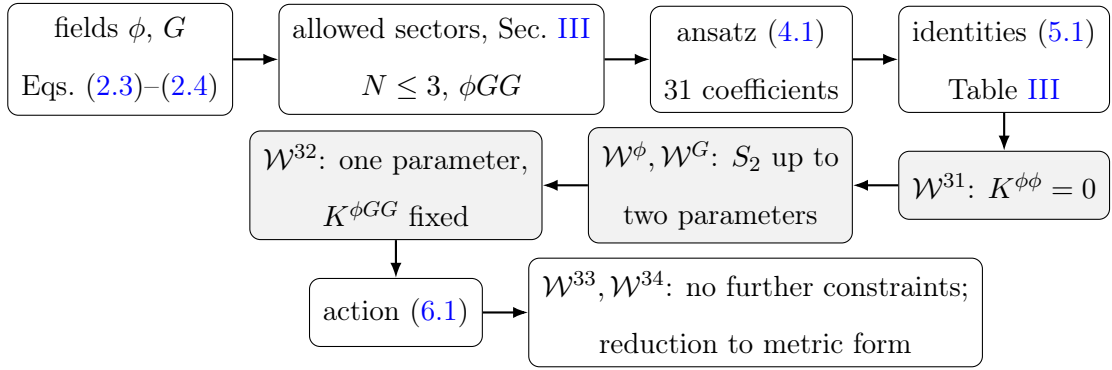

The paper is organized so that the main text carries the logic of the construction and the
appendices carry its details. Section~\ref{sec:variables} fixes the fields and their
transformations, Section~\ref{sec:closure} determines the sectors allowed in a polynomial action,
Section~\ref{sec:ansatz} gives the ansatz and its tensor bases, Section~\ref{sec:identities}
the identities and their decomposition into sectors, Section~\ref{sec:solution} the solution,
and Section~\ref{sec:checks} the reduction to the metric formulation. Figure~\ref{fig:flow} summarizes the procedure. The notation and the derivation of the
transformation laws are collected in Appendix~\ref{app:transf}; explicit tensors, bases,
identities and kernels in Appendices~\ref{app:momentum}--\ref{app:kernels}; and
Appendix~\ref{app:implementation} describes the symbolic implementation and how to reproduce
every check.

\section{First-order variables and gauge transformations}
\label{sec:variables}

\paragraph{Conventions.}
\begin{itemize}
\item Units are $\hbar=c=1$, so that dimensions are mass dimensions; the objects used
throughout and their dimensions are listed in Table~\ref{tab:notation} of
Appendix~\ref{app:transf}.
\item Spacetime has $d$ dimensions and flat metric $\eta_{\mu\nu}$ of signature
$(+,-,\dots,-)$. The algebraic tensor identities used below are polynomial in $\eta$ and the
momenta and do not depend on the signature; for generic $d$, $\sqrt{-g}$ is read as
$\sqrt{|g|}$.
\item Index symmetrization is not normalized: $h^{(\mu\nu)}\equiv h^{\mu\nu}+h^{\nu\mu}$.
\item Fourier transforms are $\p^{\mu\nu}(p)=\int d^dx\,e^{ip\cdot x}\phi^{\mu\nu}(x)$, and
likewise for $\G^{\lambda}{}_{\mu\nu}$ and for the gauge parameter $\tomega^\mu$, with
$\ip{p}\equiv\int d^dp/(2\pi)^d$. A derivative acting on a field of momentum $p$ becomes
$-ip_\mu$, and a product of fields becomes a convolution.
\item The gravitational coupling $\kappa$ satisfies $\kappa^2\propto G_N$ (Newton's gravitational constant), so that
$[\kappa]=(2-d)/2$; its normalization is fixed in Section~\ref{sec:solution}, where the
solution is identified with $\kappa^{-2}\int d^dx\,h^{\mu\nu}R_{\mu\nu}(\Gamma)$.
\end{itemize}

\paragraph{Compact notation.} In the main text indices are suppressed and a field carries
its momentum as a subscript: $\phi_p$, $G_p$ and $\omega_p$ stand for $\p^{\mu\nu}(p)$,
$\G^{\lambda}{}_{\mu\nu}(p)$ and $\tomega^\mu(p)$. A kernel $K(A,B)$ or $K(A,B,C)$ denotes the
full contraction of each argument with the corresponding index group of $K$, for instance
$K^{\phi G}(\phi_{-p},G_p)\equiv\p^{\mu\nu}(-p)\,K^{\phi G}_{\mu\nu\lambda}{}^{\rho\sigma}(p)\,\G^{\lambda}{}_{\rho\sigma}(p)$.
The field-independent and field-linear parts of the transformations are written as
$D_{\phi^0}(p)\,\omega_p$, linear in $\omega$, and $D_{\phi^1}(p,q)(\phi_{p-q},\omega_q)$, bilinear in the field and $\omega$, and
likewise for $G$; the label names the field produced and the order in $\kappa$
(Eq.~\eqref{eq:dmom} below and Appendix~\ref{app:momentum}).

\paragraph{Fields.} The independent fields are the perturbation of the Goldberg
density~\cite{Goldberg1958} and
the trace-adjusted connection, both normalized with the gravitational coupling $\kappa$,
\begin{equation}
\kappa\,\phi^{\mu\nu}=\sqrt{-g}\,g^{\mu\nu}-\eta^{\mu\nu},\qquad
\kappa\,G^{\lambda}{}_{\mu\nu}=\Gamma^{\lambda}{}_{\mu\nu}
-\tfrac12\big(\delta^{\lambda}_{\mu}\Gamma^{\sigma}{}_{\nu\sigma}
+\delta^{\lambda}_{\nu}\Gamma^{\sigma}{}_{\mu\sigma}\big),
\label{eq:Gdef}
\end{equation}
both real and symmetric in $\mu\nu$. Here $\Gamma^{\lambda}{}_{\mu\nu}$ is an independent
symmetric affine connection, the Palatini variable~\cite{Palatini1919,Ferraris1982}, and not
the Levi-Civita connection built
from $g_{\mu\nu}$: for the action found below, the field equation of $\Gamma$ identifies the
two when $d>2$ and the metric is nondegenerate, and that equation is not imposed in the
construction. The second relation in Eq.~\eqref{eq:Gdef} is a linear change of variables,
invertible for $d\neq1$. Taking its trace
gives $\Gamma^{\sigma}{}_{\nu\sigma}=\frac{2\kappa}{1-d}\,g_\nu$, with
$g_\nu\equiv G^{\sigma}{}_{\nu\sigma}$, and substituting back,
\begin{equation}
\Gamma^{\lambda}{}_{\mu\nu}=\kappa\Big[G^{\lambda}{}_{\mu\nu}
+\frac{1}{1-d}\big(\delta^{\lambda}_{\mu}\,g_\nu+\delta^{\lambda}_{\nu}\,g_\mu\big)\Big],
\label{eq:GammaofG}
\end{equation}
Eqs.~(6)--(8) of Ref.~\cite{BrandtMcKeon2015}. The trace adjustment is what cancels the
$\partial_\nu\Gamma^{\lambda}{}_{\mu\lambda}$ term of $h^{\mu\nu}R_{\mu\nu}(\Gamma)$ against
the trace part of $\partial_\lambda\Gamma^{\lambda}{}_{\mu\nu}$, leaving the single derivative
term of Eq.~\eqref{eq:result} below. Like $\phi$, which is the metric-density perturbation divided by
$\kappa$ and has mass dimension $(d-2)/2$, $G$ is the trace-adjusted connection divided by
$\kappa$ and has mass dimension $d/2$, instead of the dimension $1$ of a connection. The definitions in
Eqs.~\eqref{eq:Gdef}--\eqref{eq:GammaofG} serve only to fix the transformation laws below;
from then on $\phi$ and $G$ are treated as independent fields, and no relation between $G$ and
a metric is assumed.

\paragraph{Transformations.} Under an infinitesimal diffeomorphism
$x^\mu\to x^\mu-\kappa\,\omega^\mu(x)$ the fields transform as
$\delta\phi=\delta^{(0)}\phi+\kappa\,\delta^{(1)}\phi$ and $\delta G=\delta^{(0)}G+\kappa\,\delta^{(1)}G$, with
\begin{align}
\delta^{(0)}\phi^{\mu\nu}&=\eta^{\mu\nu}\partial\!\cdot\!\omega-\partial^{(\mu}\omega^{\nu)},&
\delta^{(1)}\phi^{\mu\nu}&=\partial_\rho(\phi^{\mu\nu}\omega^\rho)-\partial_\rho\omega^{(\mu}\phi^{\nu)\rho},
\label{eq:dphi}\\
\delta^{(0)}G^{\lambda}{}_{\mu\nu}&=\partial_\mu\partial_\nu\omega^\lambda-\tfrac12\delta^{\lambda}_{(\mu}\partial_{\nu)}\partial\!\cdot\!\omega,&
\delta^{(1)}G^{\lambda}{}_{\mu\nu}&=\omega^\rho\partial_\rho G^{\lambda}{}_{\mu\nu}-G^{\rho}{}_{\mu\nu}\partial_\rho\omega^\lambda
+G^{\lambda}{}_{\rho(\mu}\partial_{\nu)}\omega^\rho.
\label{eq:dG}
\end{align}
The split into a field-independent contribution $\delta^{(0)}$ and a field-linear contribution
$\kappa\delta^{(1)}$ is what organizes the whole construction.
Eq.~\eqref{eq:dphi} follows from the transformation of $h^{\mu\nu}=\eta^{\mu\nu}+\kappa\phi^{\mu\nu}$
as a tensor density of weight one, with the background $\eta^{\mu\nu}$ held fixed;
Eq.~\eqref{eq:dG} follows from the transformation of an affine connection through
Eq.~\eqref{eq:Gdef}. Both are derived in Appendix~\ref{app:transf}; they are
Eqs.~(17b) and (18) of Ref.~\cite{BrandtMcKeon2015}. These two laws are the only input of the
construction beyond the class of actions defined in the next section.

In momentum space the derivatives become momenta and $\delta^{(1)}$ becomes a convolution:
\begin{equation}
\delta\phi_p=D_{\phi^0}(p)\,\omega_p+\kappa\ip{q}D_{\phi^1}(p,q)\,(\phi_{p-q},\omega_q),\qquad
\delta G_p=D_{G^0}(p)\,\omega_p+\kappa\ip{q}D_{G^1}(p,q)\,(G_{p-q},\omega_q).
\label{eq:dmom}
\end{equation}
The four tensors $D_{\phi^0},D_{\phi^1},D_{G^0},D_{G^1}$ are listed in Appendix~\ref{app:momentum}.

\section{Allowed sectors of a polynomial action}
\label{sec:closure}

Before the kernels are specified, the transformation laws already restrict which field
structures a polynomial action can contain. The argument uses
Eqs.~\eqref{eq:dphi}--\eqref{eq:dG} only for a particular choice of $\omega^\mu$.

\paragraph{Coupling assignment and derivative counting.} Since $\kappa\omega$ is a
coordinate displacement, $[\kappa\omega]=-1$, and Eqs.~\eqref{eq:dphi}--\eqref{eq:dG} give
$[\phi]=(d-2)/2$ and $[G]=d/2$, with $[\kappa]=(2-d)/2$. We consider local actions
\begin{equation}
S=\sum_{n=2}^{N}\kappa^{n-2}S_n ,
\label{eq:polyaction}
\end{equation}
where $S_n$ is homogeneous of degree $n$ in the fields, Lorentz covariant and parity even,
built with $\eta$ and derivatives, and contains no dimensionful constant; $N$ is finite and
is the highest degree for which $S_N\neq0$ as a functional, that is, up to total
derivatives. The
power $\kappa^{n-2}$ and the absence of any other dimensionful constant are both needed:
additional powers of $\kappa$ at fixed $n$, or a new dimensionful parameter, would allow
terms with more derivatives and the correct dimension.\footnote{Equivalently, the action is
$\kappa^{-2}$ times a functional of $h-\eta=\kappa\phi$ and $\kappa G$ that is independent of
$\kappa$ and contains no other dimensionful constant; a term of degree $n$ in $\phi,G$ then
carries $\kappa^{n-2}$.}
A local monomial $\mathcal L_{a,b,r}$ with $a$ fields $\phi$, $b$ fields $G$ and $r$
derivatives, multiplied by $\kappa^{a+b-2}$, has mass dimension $d+b+r-2$, so that a
dimensionless action requires
\begin{equation}
b+r=2 .
\label{eq:counting}
\end{equation}
Thus $\phi\phi$ carries two derivatives, $\phi G$ one and $GG$ none; at cubic order
$\phi\phi\phi$ carries two, $\phi\phi G$ one and $\phi GG$ none, and $GGG$ is excluded. The
counting does not limit the number of fields $\phi$.

\paragraph{The highest-order identity.} The transformation $\delta=\delta^{(0)}+\kappa\delta^{(1)}$
is exactly affine in the fields: $\delta^{(0)}$ lowers the degree by one and $\delta^{(1)}$
preserves it. Collecting powers of $\kappa$ in $\delta S=0$ gives
\begin{equation}
\delta^{(0)}S_2=0,\qquad
\delta^{(0)}S_{n+1}+\delta^{(1)}S_n=0\quad(2\le n\le N-1),\qquad
\delta^{(1)}S_N=0 .
\label{eq:hierarchy}
\end{equation}
The last identity has no compensating term, because there is no $S_{N+1}$. These are
identities between integrated functionals, that is, between local densities up to total
derivatives, and every statement below that a sector vanishes is meant in this sense.

\paragraph{A dilatation.} The identities hold for every $\omega^\mu$, in particular for
$\omega^\mu=\varepsilon x^\mu$, which for $|\kappa\varepsilon|\ll1$ is the dilatation
$x^\mu\to(1-\kappa\varepsilon)x^\mu$. To avoid conditions on a parameter that grows at
infinity, the fields are taken of compact support and $\omega^\mu=\varepsilon\chi(x)x^\mu$,
with $\chi$ smooth, of compact support and equal to one on a neighborhood of their supports;
the derivatives of $\chi$ then do not contribute, and integrations by parts leave no boundary
terms. This is a particular
diffeomorphism, not an additional symmetry: it gives necessary conditions, which suffice to
exclude sectors but not to construct the action. Unlike translations and Lorentz transformations,
it is not an isometry of $\eta$. For the isometries,
$\delta^{(0)}\phi=\delta^{(0)}G=0$, and the identities express the Poincar\'e invariance
already built into the action. For the dilatation, however, $\delta^{(0)}\phi\neq0$. With $\partial_\nu\omega^\mu=\varepsilon
\delta^\mu_\nu$ and $\partial_\alpha\partial_\beta\omega^\mu=0$,
Eqs.~\eqref{eq:dphi}--\eqref{eq:dG} give
\begin{equation}
\begin{aligned}
\delta^{(1)}\phi&=\varepsilon\,(x\cdot\partial+d-2)\,\phi,&\qquad
\delta^{(1)}G&=\varepsilon\,(x\cdot\partial+1)\,G,\\
\delta^{(0)}\phi^{\mu\nu}&=\varepsilon\,(d-2)\,\eta^{\mu\nu},&\qquad
\delta^{(0)}G&=0 ;
\end{aligned}
\label{eq:dilat}
\end{equation}
$\delta^{(0)}G$ vanishes because the inhomogeneous part of the transformation of a
connection involves second derivatives of $\omega$. The numbers $d-2$ and $1$ are weights
under this transformation, not mass dimensions.\footnote{The use of a rescaling to organize
the interaction sectors was suggested by earlier work on thermal gravity, in which the leading
high-temperature graviton amplitudes obey, besides the Ward identities of diffeomorphism
invariance, Weyl identities associated with rescalings of the metric at fixed
coordinates~\cite{BrandtFrenkel1993}. The transformation used here is of a different kind: a
coordinate dilatation, which is a particular diffeomorphism and requires no symmetry beyond
those already assumed.} Each derivative adds one unit to the weight,
since $\partial_\alpha$ acting on $x^\rho$ gives $\delta^\rho_\alpha$, and contractions with
the fixed background $\eta$ add nothing. For a local density $\mathcal L_{a,b,r}$ therefore
$\delta^{(1)}\mathcal L_{a,b,r}=\varepsilon\,[\,x\cdot\partial+a(d-2)+b+r\,]\,\mathcal L_{a,b,r}$,
and since $\int d^dx\,x\cdot\partial\mathcal L=-d\int d^dx\,\mathcal L$, the integrated
functional $S_{a,b,r}$ satisfies
\begin{equation}
\delta^{(1)}S_{a,b,r}=\varepsilon\,[\,a(d-2)+b+r-d\,]\,S_{a,b,r}
=\varepsilon\,(a-1)(d-2)\,S_{a,b,r},
\label{eq:weight}
\end{equation}
where Eq.~\eqref{eq:counting} was used in the last step. Table~\ref{tab:weights} lists the
values for the sectors with two and three fields.

\begin{table}[tb]
\centering
\setlength{\tabcolsep}{8pt}
\begin{tabular}{@{}lcccccc@{}}
\toprule
Sector & $\phi\phi$ & $\phi G$ & $GG$ & $\phi GG$ & $\phi\phi G$ & $\phi\phi\phi$ \\
$(a,b,r)$ & $(2,0,2)$ & $(1,1,1)$ & $(0,2,0)$ & $(1,2,0)$ & $(2,1,1)$ & $(3,0,2)$ \\
$\delta^{(1)}S_{a,b,r}/(\varepsilon S_{a,b,r})$ & $d-2$ & $0$ & $-(d-2)$ & $0$ & $d-2$ & $2(d-2)$ \\
\bottomrule
\end{tabular}
\caption{Weights of the sectors with two and three fields under the dilatation
$\omega^\mu=\varepsilon x^\mu$, Eq.~\eqref{eq:weight}.}
\label{tab:weights}
\end{table}

\paragraph{Termination at cubic order.} In the highest-order identity, Eq.~\eqref{eq:weight}
gives
\begin{equation}
\delta^{(1)}S_N=\varepsilon\,(d-2)\sum_{a+b=N}(a-1)\,S_{a,b,2-b}=0 .
\end{equation}
Sectors with different $a$ have different field content, and the identity holds for
arbitrary independent fields (rescaling $\phi\to u\phi$ and $G\to vG$ separates them), so
for $d>2$ every sector with $a\neq1$ vanishes. The surviving sector has $a=1$, hence $b=N-1$
and $r=3-N$, and $r\ge0$ requires $N\le3$. A polynomial action therefore terminates at most at
cubic order, and its cubic part can only be the derivative-free vertex $\phi GG$; the sectors
$\phi\phi G$, with a derivative on either field, and $\phi\phi\phi$ are excluded. The same
mechanism forbids termination in the metric formulation: written in the density
$h^{\mu\nu}=\eta^{\mu\nu}+\kappa\phi^{\mu\nu}$ alone, every sector has $b=0$ and $r=2$, and the
highest one, with $a=N\ge2$, has nonvanishing weight, so its variation must be compensated at
the next order and the series does not terminate~\cite{BrandtFrenkel1996}. The weights depend
on the variables: for the covariant perturbation $g_{\mu\nu}=\eta_{\mu\nu}+\kappa h_{\mu\nu}$
the dilatation gives $h_{\mu\nu}$ weight $2$, and Eq.~\eqref{eq:weight} does not apply.

\paragraph{The identity of order $\kappa$.} With $S_3=S_{1,2,0}$, the identity
$\delta^{(0)}S_3+\delta^{(1)}S_2=0$ can be evaluated in the same way. The quadratic action
$S_2=S_{2,0,2}+S_{1,1,1}+S_{0,2,0}$ contains the sectors $\phi\phi$, $\phi G$ and $GG$, with
weights $d-2$, $0$ and $-(d-2)$. Since $S_3$ is linear in $\phi$ and contains no derivatives,
Eq.~\eqref{eq:dilat} gives $\delta^{(0)}S_3=\varepsilon(d-2)\,S_3|_{\phi\to\eta}$, the
functional quadratic in $G$ obtained by replacing $\phi^{\mu\nu}$ with $\eta^{\mu\nu}$. The
identity becomes $\varepsilon(d-2)\big[S_{2,0,2}-S_{0,2,0}+S_3|_{\phi\to\eta}\big]=0$, and
since the first functional depends only on $\phi$ and the others only on $G$, for $d>2$,
\begin{equation}
S_{2,0,2}=0,\qquad S_{0,2,0}=S_3\big|_{\phi\to\eta}.
\label{eq:orderkappa}
\end{equation}
The quadratic action has no $\phi\phi$ sector, and its algebraic $GG$ term is the cubic
vertex with $\phi$ replaced by $\eta$, as when both come from a single structure
$h^{\mu\nu}(GG)_{\mu\nu}$ with $h^{\mu\nu}=\eta^{\mu\nu}+\kappa\phi^{\mu\nu}$.

\paragraph{Result.} \emph{Let $d>2$, and let $S$ be a local action of the form
\eqref{eq:polyaction}, Lorentz covariant, parity even, polynomial in $\phi^{\mu\nu}$ and
$G^{\lambda}{}_{\mu\nu}$, with no dimensionful constant other than $\kappa$, and invariant up
to total derivatives under Eqs.~\eqref{eq:dphi}--\eqref{eq:dG}. Then $N\le3$, the cubic part
contains only the derivative-free structure $\phi GG$, and the quadratic part satisfies
Eq.~\eqref{eq:orderkappa}.} The transformation laws are inputs; what is shown is that they
restrict the action. The dilatation is blind to the tensor structure, since all contractions
within a sector carry the same weight. It therefore decides which sectors may appear and
relates whole sectors to one another. The $\phi G$ sector has weight zero and is not
constrained by the dilatation, and the $\phi GG$ vertex is related only as a whole to the $GG$
term by Eq.~\eqref{eq:orderkappa}; the contractions in the $\phi G$, $GG$ and $\phi GG$ kernels
and the remaining relative coefficients are fixed only by the Noether identities with general
$\omega^\mu$, Sections~\ref{sec:identities} and~\ref{sec:solution}.

\section{The action ansatz}
\label{sec:ansatz}

By the result of Section~\ref{sec:closure}, the most general action in the class considered
is $S=S_2+\kappa S_3$ with
\begin{equation}
\begin{aligned}
S_2&=\ip{p}\Big[K^{\phi\phi}(\phi_{-p},\phi_p)+K^{\phi G}(\phi_{-p},G_p)+K^{GG}(G_{-p},G_p)\Big],\\
S_3&=\ip{p}\ip{q}K^{\phi GG}(\phi_{p+q},G_{-p},G_{-q}).
\end{aligned}
\label{eq:ansatz}
\end{equation}
By Eq.~\eqref{eq:counting}, $K^{\phi\phi}$ is quadratic and $K^{\phi G}$ linear in the
momentum $p$, while $K^{GG}$ and $K^{\phi GG}$ are momentum independent. The cubic vertex
$K^{\phi GG}$ is purely algebraic, which is the feature of the first-order formulation that
makes the determination finite. We keep the full quadratic part and do not impose
Eq.~\eqref{eq:orderkappa}: the Noether identities reproduce it
(Section~\ref{sec:solution}), which provides an independent check.

Each kernel is expanded in a basis of tensors built from
$\eta^{\mu\nu}$ and $p^\mu$, with the index symmetries inherited from the fields it
multiplies; parity-odd structures involving the Levi-Civita tensor are not considered. The
bases are generated by symmetrizing, under the index symmetries of each kernel, the tensor
monomials, that is, the products of $\eta$'s and $p$'s with the required indices and degree
in $p$; their linear independence is verified through Gram determinants
(Appendix~\ref{app:bases}). Table~\ref{tab:kernels} gives the outcome. For $d\geq4$ the
four bases are independent and contain $4+6+5+16=31$ unknown coefficients in total.

\begin{table}[htb]
\centering
\setlength{\tabcolsep}{10pt}
\begin{tabular}{@{}lp{4.6cm}lcc@{}}
\toprule
Kernel & \raggedright Index symmetries & Momentum & Monomials & Basis dimension \\
\midrule
$K^{\phi\phi}_{\mu\nu\rho\sigma}$ & \raggedright $\mu\leftrightarrow\nu$, $\rho\leftrightarrow\sigma$, $(\mu\nu)\leftrightarrow(\rho\sigma)$ & $p^2$ & $p^2\eta\eta$, $pp\eta$ & 4 \\
$K^{\phi G}_{\mu\nu\lambda\rho\sigma}$ & \raggedright $\mu\leftrightarrow\nu$, $\rho\leftrightarrow\sigma$ & $p$ & $p\eta\eta$ & 6 \\
$K^{GG}_{\lambda\mu\nu\rho\sigma\tau}$ & \raggedright $\mu\leftrightarrow\nu$, $\sigma\leftrightarrow\tau$, $(\lambda\mu\nu)\leftrightarrow(\rho\sigma\tau)$ & $p^0$ & $\eta\eta\eta$ & 5 \\
$K^{\phi GG}_{\mu\nu\lambda\alpha\beta\rho\sigma\tau}$ & \raggedright $\mu\leftrightarrow\nu$, $\alpha\leftrightarrow\beta$, $\sigma\leftrightarrow\tau$, $(\lambda\alpha\beta)\leftrightarrow(\rho\sigma\tau)$ & $p^0$ & $\eta\eta\eta\eta$ & 16 \\
\bottomrule
\end{tabular}
\caption{The four kernels of the ansatz \eqref{eq:ansatz}, their symmetries, and the
dimension of the tensor basis in which each is expanded, for $d\geq4$.}
\label{tab:kernels}
\end{table}

\section{Noether identities and their decomposition}
\label{sec:identities}

For $N=3$ the hierarchy \eqref{eq:hierarchy} consists of three identities,
\begin{equation}
\delta^{(0)}S_2=0,\qquad
\delta^{(0)}S_3+\delta^{(1)}S_2=0,\qquad
\delta^{(1)}S_3=0,
\label{eq:noether}
\end{equation}
where $\delta^{(m)}S_n\equiv\int(\delta S_n/\delta A)\,\delta^{(m)}A$ summed over the
fields $A=\phi,G$. Each identity holds for arbitrary $\omega$ and arbitrary field
configurations, so it can be decomposed further according to which fields appear. Taking
functional derivatives with respect to the fields and setting the fields to zero turns each
piece into an algebraic tensor identity among the kernels, in momentum space. The
decomposition is summarized in Table~\ref{tab:identities}; the six identities are written out
in Appendix~\ref{app:identities}. The resulting relations among the action kernels are also
called classical Ward identities, the terminology used in the accompanying notebooks; we
keep ``Noether identities'' to emphasize their derivation from Noether's second theorem.
They concern the classical action before gauge fixing, not the Green functions or the BRST
identities of a gauge-fixed theory.

\begin{table}[htb]
\centering
\setlength{\tabcolsep}{10pt}
\begin{tabular}{@{}llllc@{}}
\toprule
Identity & From & Field sector & Kernels involved & Projection basis \\
\midrule
$\W^{\phi}$  & $\delta^{(0)}S_2$ & $\phi$   & $K^{\phi\phi}$, $K^{\phi G}$ & 3 \\
$\W^{G}$     & $\delta^{(0)}S_2$ & $G$      & $K^{\phi G}$, $K^{GG}$ & 6 \\
$\W^{31}$    & $\delta^{(1)}S_2$ & $\phi\phi$ & $K^{\phi\phi}$ & 7 \\
$\W^{32}$    & $\delta^{(0)}S_3+\delta^{(1)}S_2$ & $\phi G$ & $K^{\phi GG}$, $K^{\phi G}$ & 23 \\
$\W^{33}$    & $\delta^{(0)}S_3+\delta^{(1)}S_2$ & $GG$ & $K^{\phi GG}$, $K^{GG}$ & 20 \\
$\W^{34}$    & $\delta^{(1)}S_3$ & $\phi GG$ & $K^{\phi GG}$ & 95 \\
\bottomrule
\end{tabular}
\caption{Decomposition of the Noether identities \eqref{eq:noether} into field sectors. The
last column is the generic-dimension size of the projection basis onto which each identity
is projected to extract scalar equations for the coefficients, with the momenta
set equal as described in the text; dimension-specific dependencies, such as that of the
95-element basis at $d=4$, are discussed in Appendix~\ref{app:identities}.}
\label{tab:identities}
\end{table}

The identities $\W^{\phi}$ and $\W^{G}$, which involve only $\delta^{(0)}$ and $S_2$, are
labeled by the field sector; in the others, which involve $\delta^{(1)}$ or $S_3$, the
superscript $3$ follows the accompanying notebook, where these are grouped as cubic
identities, and the second digit numbers the field sectors $\phi\phi$, $\phi G$, $GG$ and
$\phi GG$. Two remarks clarify the structure of these identities. First,
$\W^{31}$ involves the $\phi\phi$ kernel alone, because $S_3$ contains two factors of $G$ and
$\delta^{(1)}G$ is linear in $G$: nothing in the cubic sector can contribute to the $\phi\phi$ sector. This uses the absence of $\phi\phi\phi$ and $\phi\phi G$ terms, established in
Section~\ref{sec:closure}: such terms would contribute to this sector through $\delta^{(0)}\phi$
and $\delta^{(0)}G$.
Second, the cubic kernel $K^{\phi GG}$ enters $\W^{32}$ and $\W^{33}$, which relate it to
the quadratic kernels through the field-dependent part $\delta^{(1)}$ of the
transformations, and $\W^{34}$, which constrains the cubic kernel alone. Within the chosen
ansatz, these identities couple the cubic vertex to the quadratic action; the solution
below establishes that no independent cubic coefficient remains.

Each identity is turned into scalar equations by projecting it onto a basis of tensors with
the same index structure and momentum dependence; since the identity vanishes, so does every
projection. Conversely, for an identity that can be expanded in a projection basis with nonsingular
Gram matrix, the vanishing of all projections is equivalent to the vanishing of the identity;
the one exception, $\W^{34}$ at $d=4$, is discussed in Appendix~\ref{app:identities}.

\paragraph{Identified momenta.} In the identities with two or three momenta the momenta are
set equal to one another, $p_1=p_2=p_3\equiv p$, with $p$ itself arbitrary, so that every
projection basis is built from $\eta$ and a single vector; the dimensions in
Table~\ref{tab:identities} refer to these bases. $\W^{\phi}$ and $\W^{G}$ involve a single
momentum and are used as they stand. The Noether identities hold as tensor identities for
arbitrary momenta, in particular for equal ones, so the restriction to equal momenta can only weaken the
system: it yields necessary conditions, never spurious ones, and the solution space of the
equal-momentum system contains that of the full one. The equal-momentum system already has a
one-dimensional solution space (Section~\ref{sec:solution}), spanned by the
Goldberg--Palatini action. For this action the identities hold for arbitrary momenta, since
$h^{\mu\nu}R_{\mu\nu}(\Gamma)$ is a scalar density under
Eqs.~\eqref{eq:dphi}--\eqref{eq:dG} by construction and its variation is a total derivative,
which integrates to zero for gauge parameters of compact support. The two solution spaces
therefore coincide, and nothing is lost by the identification.

\section{Solution}
\label{sec:solution}

The identities are generated and the resulting linear system is solved symbolically with
FeynCalc~\cite{FeynCalc} in Mathematica~\cite{Mathematica} (Appendix~\ref{app:implementation}).
The system is solved in a sequence that follows the structure of
Table~\ref{tab:identities}, shown in Table~\ref{tab:flow}.

\begin{table}[htb]
\centering
\setlength{\tabcolsep}{10pt}
\begin{tabular}{@{}llp{5.2cm}l@{}}
\toprule
Step & Identity & \raggedright Constraints imposed & Remaining freedom \\
\midrule
1 & $\W^{31}$ & \raggedright all 4 of $K^{\phi\phi}$, to zero & 27 \\
2 & $\W^{\phi}$, $\W^{G}$ & \raggedright 9 of the 11 in $K^{\phi G}$, $K^{GG}$ & 2 in $S_2$; 16 in $K^{\phi GG}$ \\
3 & $\W^{32}$ & \raggedright the remaining relative freedom in $S_2$; all 16 of $K^{\phi GG}$ & 1 (overall constant) \\
4 & $\W^{33}$, $\W^{34}$ & \raggedright none: both are satisfied identically & 1 \\
\bottomrule
\end{tabular}
\caption{Sequential solution of the constraint system. The third column counts
independent constraints, that is, parameters eliminated at each step; which individual
coefficients are expressed in terms of which depends on the parametrization chosen.}
\label{tab:flow}
\end{table}

Step~1 shows that, within the ansatz, the identities force the purely metric kernel to vanish,
$K^{\phi\phi}=0$, in agreement with Eq.~\eqref{eq:orderkappa}, before any comparison with a known action: in this formalism the
dynamics of $\phi$ is carried entirely by its coupling to $G$. At quadratic order alone, $\W^{\phi}$ and $\W^{G}$ would allow a $K^{\phi\phi}$ of
Fierz--Pauli form, invariant by itself under $\delta^{(0)}\phi$; it would merely duplicate the
graviton kinetic term generated by the elimination of $G$, Eq.~\eqref{eq:second}, and it is the field-dependent identity $\W^{31}$
that removes it. Steps~2 and~3 then leave only
the overall constant, the cubic identity in the $\phi G$ sector fixing both the relative
coefficient left in $S_2$ and all sixteen coefficients of $K^{\phi GG}$, and Step~4 imposes
no further condition. The solution is therefore unique up to an overall constant.

Transforming back to configuration space, and writing $g_\mu\equiv G^{\sigma}{}_{\mu\sigma}$ for
the trace of $G$ and $\operatorname{tr}(G_\mu G_\nu)\equiv G^{\lambda}{}_{\mu\sigma}G^{\sigma}{}_{\nu\lambda}$
for the matrix trace over the upper and one lower index, the action selected by the identities reads
\begin{equation}
S=\int d^dx\,\Big[-G^{\lambda}{}_{\mu\nu}\partial_\lambda\phi^{\mu\nu}
+\big(\eta+\kappa\,\phi\big)^{\mu\nu}
\Big(-\operatorname{tr}(G_\mu G_\nu)+\frac{1}{d-1}\,g_\mu g_\nu\Big)\Big],
\label{eq:result}
\end{equation}
which gives the Goldberg--Palatini action
$\frac{1}{\kappa^2}\int d^dx\,h^{\mu\nu}R_{\mu\nu}(\Gamma)$, up to a surface term, written in the variables
\eqref{eq:Gdef}, with $h^{\mu\nu}=\eta^{\mu\nu}+\kappa\phi^{\mu\nu}$; Ref.~\cite{BrandtMcKeon2016}
writes the same action with $\kappa=1$, Eqs.~(3.5)--(3.6) there. With the field dimensions
and coupling-power assignment adopted here, the overall constant left undetermined by the
identities is dimensionless. Both relations of Eq.~\eqref{eq:orderkappa} can be read off
Eq.~\eqref{eq:result}: there is no $\phi\phi$ term, and since $(\eta+\kappa\phi)^{\mu\nu}$
multiplies a single structure, the algebraic $GG$ term is the cubic vertex with $\phi$
replaced by $\eta$.
The explicit momentum-space kernels are given in Appendix~\ref{app:kernels}.

\section{Reduction to the metric formulation}
\label{sec:checks}

Writing the $GG$ term of Eq.~\eqref{eq:result}
as $\tfrac12 M(\eta+\kappa\phi)(G,G)$, with
$M(h)(G,G)=2h^{\mu\nu}\big(-\operatorname{tr}(G_\mu G_\nu)+\tfrac{1}{d-1}g_\mu g_\nu\big)$
linear in $h$, the field equation of $G$ is algebraic, $M(\eta+\kappa\phi)\,G=\partial\phi$.
For $d>2$ and nondegenerate $h$ its solution, substituted back, gives
\begin{equation}
S=-\tfrac12\int d^dx\;M^{-1}(\eta+\kappa\phi)\big(\partial\phi,\partial\phi\big),
\label{eq:second}
\end{equation}
the second-order Einstein--Hilbert action in the Goldberg variable to all orders in $\phi$,
up to a surface term, Eq.~(3.13) of Ref.~\cite{BrandtMcKeon2016}. The formal expansion
$M^{-1}(\eta+\kappa\phi)=M^{-1}(\eta)-\kappa\,M^{-1}(\eta)M(\phi)M^{-1}(\eta)+\dots$,
Eq.~(3.18) there, generates the quadratic action and every $n$-point vertex; with
$G_0\equiv M^{-1}(\eta)\,\partial\phi$, the cubic one is
\begin{equation}
S^{(3)}=\frac{\kappa}{2}\int d^dx\;M(\phi)\big(G_0,G_0\big).
\label{eq:cubicvertex}
\end{equation}

The reduction has been checked symbolically at quadratic and cubic order against the
Lagrangians of Capper, Leibbrandt and Ram\'on Medrano (CLM)~\cite{CLM1973}, Eqs.~(2.10) and
(2.16) of that reference, which are written in the same metric-density perturbation
$\sqrt{-g}\,g^{\mu\nu}=\eta^{\mu\nu}+\kappa\phi^{\mu\nu}$. Starting from the bootstrap
kernels, the accompanying notebook, written with xAct~\cite{xAct,xPerm,xTras}
(Appendix~\ref{app:implementation}), derives
$M_0^{-1}$, with $M_0\equiv M(\eta)$, verifies the left and right inverse relations,
constructs $G_0=M_0^{-1}\partial\phi$, and evaluates the quadratic Lagrangian from
Eq.~\eqref{eq:second} and the cubic one from Eq.~\eqref{eq:cubicvertex}. The conventions
must be kept explicit: CLM write $2K^{-2}\int d^dx\,h^{\mu\nu}R^{\rm CLM}_{\mu\nu}$ with
$R^{\rm CLM}_{\mu\nu}=-R_{\mu\nu}$ in our convention (Eqs.~(2.1) and (2.2) of
Ref.~\cite{CLM1973}), whereas Eq.~\eqref{eq:result}
corresponds to $\kappa^{-2}\int d^dx\,h^{\mu\nu}R_{\mu\nu}$ up to a surface term. With
$K=\kappa$ and $\mathcal L^{\rm CLM}_{(j)}$ the coefficients of their expansion
$\mathcal L=\sum_{j\ge2}K^{j-2}\mathcal L^{\rm CLM}_{(j)}$, the symbolic checks give
\begin{subequations}
\label{eq:clm}
\begin{align}
\mathcal L^{(2)}+\tfrac12\,\mathcal L^{\rm CLM}_{(2)}&=\partial_\mu J^\mu,\qquad
J^\mu=\tfrac12\big(\phi^{\nu\rho}\partial_\rho\phi^{\mu}{}_{\nu}-\phi^{\mu\nu}\partial_\rho\phi^{\rho}{}_{\nu}\big),
\label{eq:clm-quadratic}\\
\mathcal L^{(3)}+\tfrac{\kappa}{2}\,\mathcal L^{\rm CLM}_{(3)}&=0,
\label{eq:clm-cubic}
\end{align}
\end{subequations}
with indices raised and lowered with $\eta$.\footnote{In transcribing Eq.~(2.16) of
Ref.~\cite{CLM1973}, the first term of its second group is read as
$\phi_{\mu\tau}\phi_{\mu\sigma,\lambda}\phi_{\lambda\tau,\sigma}$, which restores consistent
index contractions in the printed expression.} The cubic equality is local and requires
neither integration by parts nor the field equations of $\phi$; the quadratic actions agree
under boundary conditions that remove the integral of $\partial_\mu J^\mu$. The surface term reflects the
form in which CLM write their quadratic Lagrangian: expressed in the full density
$h^{\mu\nu}$, the action obtained by eliminating $G$ from Eq.~\eqref{eq:result} coincides
identically with Eq.~(A14) of Ref.~\cite{CLM1973}, after accounting for the relative
normalization stated above and without integration by parts, so that their
Eq.~(2.10) differs from the expansion of their Eq.~(A14) by a total derivative. The corresponding
three-graviton vertices therefore agree once the action, Fourier and symmetrization
conventions are matched, without a separate transcription of CLM's momentum-space vertex.

\paragraph*{Range of validity.} Uniqueness holds under the assumptions of the Result of
Section~\ref{sec:closure}, which do not include a restriction of the cubic sector, and
for $d\geq4$, where the 31 basis elements are independent; at $d=3$ the cubic basis
degenerates and the count of Section~\ref{sec:ansatz} does not apply, although the
result of Section~\ref{sec:closure} itself holds for $d>2$. It is not established for nonpolynomial actions, for additional powers of $\kappa$ or independent
dimensionful couplings, which allow sectors with $b+r=4$ such as quadratic curvature terms,
or for $d=2$. The overall
constant is not fixed by the homogeneous invariance conditions; it requires an external
normalization, such as matching the weak-field equations to the Newtonian limit with a
specified normalization of the matter source. The construction is classical throughout;
whether the first- and second-order formulations remain equivalent beyond tree level, a
question studied in Ref.~\cite{BrandtMcKeon2016}, is not addressed here.

\section{Conclusions}
\label{sec:conclusions}

The procedure described here identifies the Goldberg--Palatini action through the
Noether identities of its prescribed field transformations, in two steps. First, a
dilatation, which is a particular diffeomorphism, shows that a local action polynomial in
$\phi$ and $G$, with the coupling-power assignment and no other dimensionful constant,
terminates at most at cubic order and that its cubic part can only be a derivative-free $\phi GG$
vertex; the restriction of the cubic sector is therefore a consequence of polynomiality
rather than an assumption. The dilatation is blind to the tensor structure of these sectors, which is fixed in the second step.
Second, within the resulting class the identities with general
$\omega^\mu$ reduce the 31 coefficients of the kernel bases (for $d\geq4$) to a single
overall constant. In particular, they eliminate the quadratic $\phi\phi$ kernel, in
agreement with the dilatation argument, and fix the relative coefficients of the mixed
$\phi G$ term, the algebraic $GG$ term and the cubic $\phi GG$ interaction. The resulting
structure is that of first-order Einstein gravity.

Individual identities can leave components of a kernel undetermined; uniqueness here follows
from imposing all identities simultaneously:
the field-dependent transformations link the quadratic and cubic sectors and remove the
freedom left by the quadratic identities alone, and the resulting homogeneous linear system
has a one-dimensional solution space. Uniqueness is thus a property of this restricted
system, not a general consequence of Noether identities. The connection is not a field
exempt from the procedure: its variation enters every sector of the identity except the
$\phi\phi$ one, and once the preceding constraints are imposed the cubic vertex is fixed by
the sector that couples the two fields, $\W^{32}$.

Since the equation for $G$ is algebraic and uniquely solvable for $d>2$ and nondegenerate
metric density, a construction that terminates at cubic order in the first-order variables
determines, through Eq.~\eqref{eq:second}, the infinite set of vertices of the metric
formulation; once the conventions are matched as in Section~\ref{sec:checks}, its quadratic
and cubic terms agree with those of Ref.~\cite{CLM1973}, the quadratic ones up to a surface
term. This is the practical content of the classical equivalence between the two
formulations, and it is what makes the first-order variables the natural setting for the
determination.

The method makes the role of symmetry explicit at each order in the fields. It avoids
selecting an energy--momentum tensor as the source of successive self-couplings and instead
imposes off-shell relations directly on the action kernels. Its practical advantage is that
the polynomial first-order formulation turns the determination of the action into a finite tensor-algebra
problem.

A natural extension is to relax the coupling-power assignment. Additional powers of
$\kappa$, or an independent dimensionful coupling, admit sectors with $b+r=4$, such as those
produced by quadratic curvature terms; there $G$ may carry derivatives, so that its auxiliary
character has to be reexamined (in Palatini $f(R)$ theories, for instance, the connection
remains nondynamical), and the counting of Section~\ref{sec:closure} has to be redone. Such
higher-derivative gravitational actions would allow the method to probe a broader space of
gauge-invariant dynamics. The case $d=2$, in which every weight in Eq.~\eqref{eq:weight}
vanishes, is also outside the argument and would require a separate treatment. For Yang--Mills theory the tree-amplitude version of the argument
is known to close and to fix the interaction structure under its stated
assumptions~\cite{BrandtFrenkel1996}; the action-level version with tensor bases would
provide a direct check of the method.

\begin{acknowledgments}
F.T.B. thanks Giacomo Farina Jolvino and Gustavo Alves de Menezes for discussions and
acknowledges partial support from CNPq (grant No.~309411/2022-0).
A.C.B.C. acknowledges an undergraduate research scholarship from the Programa Unificado de
Bolsas de Estudos of the University of S\~ao Paulo (PUB-USP).
In preparing this manuscript, AI tools were used as assistants for literature search,
language and LaTeX editing, and coding.
\end{acknowledgments}

\appendix

\section{Notation and transformation laws}
\label{app:transf}

Table~\ref{tab:notation} lists the objects used in the main text, with their index
structure, momentum dependence and mass dimension.

\begin{table}[htb]
\centering\scriptsize
\setlength{\tabcolsep}{9pt}
\begin{tabular}{@{}lllcc@{}}
\toprule
Symbol & Meaning & Indices & Momenta & Dimension \\
\midrule
$\kappa$ & gravitational coupling & & & $(2-d)/2$ \\
$\phi^{\mu\nu}$; $\phi_p$ & density perturbation, Eq.~\eqref{eq:Gdef} & $\mu\nu$ symmetric & one & $(d-2)/2$; $-(d+2)/2$ \\
$G^{\lambda}{}_{\mu\nu}$; $G_p$ & trace-adjusted connection, Eq.~\eqref{eq:Gdef} & $\mu\nu$ symmetric & one & $d/2$; $-d/2$ \\
$\omega^{\mu}$; $\omega_p$ & gauge parameter & vector & one & $(d-4)/2$; $-(d+4)/2$ \\
$D_{\phi^0}(p)$, $D_{G^0}(p)$ & $\delta^{(0)}\phi$, $\delta^{(0)}G$ acting on $\omega$ & App.~\ref{app:momentum} & $p$ & $1$, $2$ \\
$D_{\phi^1}(p,q)$, $D_{G^1}(p,q)$ & $\delta^{(1)}\phi$, $\delta^{(1)}G$ acting on (field,\,$\omega$) & App.~\ref{app:momentum} & $p,q$ & $1$, $1$ \\
$K^{\phi\phi}$, $K^{\phi G}$, $K^{GG}$ & quadratic kernels of $S_2$ & Table~\ref{tab:kernels} & $p,p,\text{none}$ & $2$, $1$, $0$ \\
$K^{\phi GG}$ & cubic kernel of $S_3$ & Table~\ref{tab:kernels} & none & $0$ \\
$\W^{\phi},\W^{G},\W^{31},\dots,\W^{34}$ & the six Noether identities & App.~\ref{app:identities} & $p$; $p_i$ & \\
\bottomrule
\end{tabular}
\caption{Notation and mass dimensions. For the fields and the gauge parameter the two entries
in the last column refer to configuration and momentum space; $[\ip{p}]=d$.}
\label{tab:notation}
\end{table}

The transformation laws \eqref{eq:dphi} and \eqref{eq:dG} are Eqs.~(17b) and (18) of
Ref.~\cite{BrandtMcKeon2015}, where $\kappa=1$ and the parameter $\theta$ is $-\kappa\omega$
here. We sketch their derivation. Under $x'^\mu=x^\mu-\varepsilon^\mu(x)$ the variation
$\delta T(x)\equiv T'(x)-T(x)$ of a tensor is its Lie derivative $\mathcal{L}_\varepsilon T$;
for a $(1,2)$ tensor,
\begin{equation}
\mathcal{L}_\varepsilon X^{\lambda}{}_{\mu\nu}=\varepsilon^\rho\partial_\rho X^{\lambda}{}_{\mu\nu}
-X^{\rho}{}_{\mu\nu}\partial_\rho\varepsilon^\lambda
+X^{\lambda}{}_{\rho\nu}\partial_\mu\varepsilon^\rho
+X^{\lambda}{}_{\mu\rho}\partial_\nu\varepsilon^\rho ,
\label{eq:lie12}
\end{equation}
and $\mathcal{L}_\varepsilon$ annihilates the Kronecker delta.

\paragraph{The metric density.} $h^{\mu\nu}\equiv\sqrt{-g}\,g^{\mu\nu}$ is a contravariant
tensor density of weight one,
\begin{equation}
\delta h^{\mu\nu}=\partial_\rho\big(h^{\mu\nu}\varepsilon^\rho\big)
-h^{\rho\nu}\partial_\rho\varepsilon^\mu-h^{\mu\rho}\partial_\rho\varepsilon^\nu ,
\label{eq:dh}
\end{equation}
which is Eq.~(17a) of Ref.~\cite{BrandtMcKeon2015}. Holding the background $\eta^{\mu\nu}$
fixed, so that $\delta h^{\mu\nu}=\kappa\,\delta\phi^{\mu\nu}$, inserting
$h^{\mu\nu}=\eta^{\mu\nu}+\kappa\phi^{\mu\nu}$ and $\varepsilon=\kappa\omega$, and dividing by
$\kappa$ gives Eq.~\eqref{eq:dphi}.

\paragraph{The connection.} An affine connection transforms as
\begin{equation}
\delta\Gamma^{\lambda}{}_{\mu\nu}=\mathcal{L}_\varepsilon\Gamma^{\lambda}{}_{\mu\nu}
+\partial_\mu\partial_\nu\varepsilon^\lambda ,
\label{eq:dGamma}
\end{equation}
where, here and below, $\mathcal{L}_\varepsilon$ applied to a non-tensorial object denotes
only the tensor-form expression \eqref{eq:lie12}. Contracting $\lambda$ with $\mu$, the
trace $T_\nu\equiv\Gamma^{\lambda}{}_{\nu\lambda}$ obeys
\begin{equation}
\delta T_\nu=\mathcal{L}_\varepsilon T_\nu+\partial_\nu(\partial\!\cdot\!\varepsilon).
\label{eq:dT}
\end{equation}
By Eq.~\eqref{eq:Gdef},
$\delta(\kappa G^{\lambda}{}_{\mu\nu})=\delta\Gamma^{\lambda}{}_{\mu\nu}
-\tfrac12\big(\delta^\lambda_\mu\,\delta T_\nu+\delta^\lambda_\nu\,\delta T_\mu\big)$. Since
$\mathcal{L}_\varepsilon$ annihilates $\delta^\lambda_\mu$, the tensor-form terms combine into
$\kappa\,\mathcal{L}_\varepsilon G$, and only the inhomogeneous part of Eq.~\eqref{eq:dT}
survives from the trace correction:
\begin{equation}
\delta(\kappa G^{\lambda}{}_{\mu\nu})=\kappa\,\mathcal{L}_\varepsilon G^{\lambda}{}_{\mu\nu}
+\partial_\mu\partial_\nu\varepsilon^\lambda
-\tfrac12\big(\delta^\lambda_\mu\partial_\nu+\delta^\lambda_\nu\partial_\mu\big)\partial\!\cdot\!\varepsilon ,
\end{equation}
which with $\varepsilon=\kappa\omega$ is Eq.~\eqref{eq:dG}. Its trace,
$\delta^{(0)}G^{\lambda}{}_{\nu\lambda}=\tfrac{1-d}{2}\,\partial_\nu\partial\!\cdot\!\omega$,
is consistent with $\Gamma^{\sigma}{}_{\nu\sigma}=\frac{2\kappa}{1-d}g_\nu$. Both
Eq.~\eqref{eq:dT} and the final result were verified symbolically
(Appendix~\ref{app:implementation}).

\section{Transformations in momentum space}
\label{app:momentum}

With the Fourier conventions of Section~\ref{sec:variables}, Eq.~\eqref{eq:dmom} reads, with
all indices displayed,
\begin{align}
\delta\p^{\mu\nu}(p)&=D_{\phi^0}{}^{\mu\nu}{}_{\alpha}(p)\,\tomega^\alpha(p)
+\kappa\ip{q}\p^{\alpha\beta}(p-q)\,\tomega^\rho(q)\,D_{\phi^1}{}^{\mu\nu}{}_{\alpha\beta\rho}(p,q),\\
\delta\G^{\lambda}{}_{\mu\nu}(p)&=D_{G^0}{}^{\lambda}{}_{\mu\nu\alpha}(p)\,\tomega^\alpha(p)
+\kappa\ip{q}\G^{\alpha}{}_{\beta\gamma}(p-q)\,\tomega^\rho(q)\,
D_{G^1}{}^{\lambda}{}_{\mu\nu\alpha}{}^{\beta\gamma}{}_{\rho}(p,q),
\end{align}
with
\begin{align}
D_{\phi^0}{}^{\mu\nu}{}_{\alpha}(p)&=-i\big(\eta^{\mu\nu}p_\alpha-p^{(\mu}\delta^{\nu)}_{\alpha}\big),\\
D_{\phi^1}{}^{\mu\nu}{}_{\alpha\beta\rho}(p,q)&=-i\big(\delta^{\mu}_{\alpha}\delta^{\nu}_{\beta}\,p_\rho
-q_\beta\,\delta^{(\mu}_{\alpha}\delta^{\nu)}_{\rho}\big),\\
D_{G^0}{}^{\lambda}{}_{\mu\nu\alpha}(p)&=-p_\mu p_\nu\delta^{\lambda}_{\alpha}
+\tfrac12\delta^{\lambda}_{(\mu}p_{\nu)}p_\alpha,\\
D_{G^1}{}^{\lambda}{}_{\mu\nu\alpha}{}^{\beta\gamma}{}_{\rho}(p,q)&=-i\big[
\delta^{\lambda}_{\alpha}\delta^{\beta}_{\mu}\delta^{\gamma}_{\nu}\,(p-q)_\rho
-q_\alpha\,\delta^{\beta}_{\mu}\delta^{\gamma}_{\nu}\delta^{\lambda}_{\rho}
+\delta^{\lambda}_{\alpha}\delta^{\beta}_{\rho}\delta^{\gamma}_{(\mu}q_{\nu)}\big].
\end{align}
These expressions are understood as acting on fields with their stated index
symmetries; symmetrization in the contracted field indices is therefore implicit, and a
representative differing by terms that vanish on symmetric fields gives the same result.

\section{Tensor bases}
\label{app:bases}

Each kernel is expanded in the basis of all independent tensors with its index structure,
built from $\eta$ and $p$ with the momentum dependence of Table~\ref{tab:kernels}. The
construction is the same for the four kernels. One enumerates the tensor monomials of the
required degree in $p$, each a single product of $\eta$'s over the index
positions of $K$ not carrying a
momentum, times the appropriate powers of $p$, and symmetrizes each monomial under the
index symmetries of the kernel. These symmetries, listed in Table~\ref{tab:kernels}, generate
a finite group $\mathsf{G}_K$ of permutations of the index positions of $K$. Applying every
element of $\mathsf{G}_K$ to a monomial $T$ produces a set of $n$ distinct monomials
$T_1,\dots,T_n$, the \emph{orbit} of $T$; the symmetrized monomial is the average over the
orbit, $\mathcal{S}[T]=\frac1n\sum_{k=1}^{n}T_k$ (equivalently, the average of $T$ over the
group). Two monomials give the same $\mathcal{S}[T]$ if and only if they lie in the same
orbit, so the orbit averages form a basis of the invariant polynomials in $\eta$ and $p$, in
one-to-one correspondence with the orbits, and any element of an orbit may serve as its
representative. This is a candidate tensor basis; its independence as a set of tensors in a
given dimension is checked below. The orbit sizes $n$ add up to the number of monomials. In
the implementation the monomials are generated by applying all permutations of the index
positions to a few prototensors (the distinct ways of distributing $\eta$'s and momenta) and
removing duplicates; the symmetrized elements are then obtained by averaging over the
kernel symmetries and removing duplicates again, which yields exactly the orbit averages
described above.

Tables~\ref{tab:basis-phiphi}--\ref{tab:basis-phiGG} list, for each kernel, a representative
monomial of every orbit and the number $n$ of terms in the corresponding symmetrized sum.
Indices are written in the order of the kernel's index positions; $p^2\equiv p_\alpha p^\alpha$. For
$K^{\phi\phi}$, for instance, element~4 of Table~\ref{tab:basis-phiphi} is
$\tfrac14\big(p_\mu p_\rho\eta_{\nu\sigma}+p_\nu p_\rho\eta_{\mu\sigma}
+p_\mu p_\sigma\eta_{\nu\rho}+p_\nu p_\sigma\eta_{\mu\rho}\big)$.

Distinct orbits have disjoint monomial content, so the elements are linearly independent as
polynomials in $\eta$ and $p$, before any dimension-dependent identity is used. They can nevertheless become dependent as tensors in low dimension,
through identities that hold only for specific $d$. Independence for generic $d$ is verified by
computing the Gram matrix $\Gamma_K$ of each basis, $(\Gamma_K)_{ij}=\mathcal{S}[T_i]\cdot\mathcal{S}[T_j]$
with all indices contracted. Its determinant is a polynomial in $d$ and $p^2$: when it is
nonzero the basis is linearly independent, in any signature, and its zeros locate the
dimensions where the basis degenerates (at Euclidean momentum the contraction pairing is
positive definite, so there a vanishing determinant implies a dependence). The bases used
here depend on a single momentum and their elements share the same degree in $p$, so a
dependence with coefficients depending on $p^2$ is equivalent to one with constant
coefficients; with several independent momenta, a vanishing Gram determinant need not imply
a dependence with constant coefficients. For the bases of Tables~\ref{tab:basis-phiphi}--\ref{tab:basis-phiGG}, up to
positive numerical factors,
\begin{align}
\det\Gamma_{\phi\phi}&=(p^2)^8\,(d-2)(d-1)^3(d+2)(d+4),\\
\det\Gamma_{\phi G}&=(p^2)^6\,(d-2)(d-1)^5(d+2)^5(d+4),\\
\det\Gamma_{GG}&=d^5\,(d-2)(d-1)^4(d+2)^4(d+4),\\
\det\Gamma_{\phi GG}&=d^{16}\,(d-3)(d-2)^7(d-1)^{15}(d+1)^3(d+2)^{15}(d+4)^6(d+6).
\end{align}
All four are nonzero for $d\geq4$ and $p^2\neq0$; the powers of $p^2$ reflect the momentum
dependence of the first two bases and are unrelated to dimension-dependent relations. The
zeros at $d=1,2$, and at $d=3$ for $K^{\phi GG}$,
are the low dimensions in which products of $\eta$'s with many indices become linearly
dependent (in $d=3$, for instance, antisymmetrization over four indices vanishes).

\begin{table}[htb]
\centering
\begin{tabular}{@{}cl c@{}}
\toprule
& Representative monomial & $n$ \\
\midrule
1 & $p^2\,\eta_{\mu\nu}\eta_{\rho\sigma}$ & 1 \\
2 & $p^2\,\eta_{\mu\rho}\eta_{\nu\sigma}$ & 2 \\
3 & $p_{\mu}p_{\nu}\eta_{\rho\sigma}$ & 2 \\
4 & $p_{\mu}p_{\rho}\eta_{\nu\sigma}$ & 4 \\
\bottomrule
\end{tabular}
\caption{Basis of $K^{\phi\phi}_{\mu\nu\rho\sigma}(p)$: 4 elements, the orbits of the 9 monomials
under the symmetry group of order 8.}
\label{tab:basis-phiphi}
\end{table}

\begin{table}[htb]
\centering
\begin{tabular}{@{}cl c@{}}
\toprule
& Representative monomial & $n$ \\
\midrule
1 & $p_{\mu}\eta_{\nu\lambda}\eta_{\rho\sigma}$ & 2 \\
2 & $p_{\mu}\eta_{\nu\rho}\eta_{\lambda\sigma}$ & 4 \\
3 & $p_{\lambda}\eta_{\mu\nu}\eta_{\rho\sigma}$ & 1 \\
4 & $p_{\lambda}\eta_{\mu\rho}\eta_{\nu\sigma}$ & 2 \\
5 & $p_{\rho}\eta_{\mu\nu}\eta_{\lambda\sigma}$ & 2 \\
6 & $p_{\rho}\eta_{\mu\lambda}\eta_{\nu\sigma}$ & 4 \\
\bottomrule
\end{tabular}
\caption{Basis of $K^{\phi G}_{\mu\nu\lambda\rho\sigma}(p)$: 6 elements, the orbits of the 15
monomials under the symmetry group of order 4.}
\label{tab:basis-phiG}
\end{table}

\begin{table}[htb]
\centering
\begin{tabular}{@{}cl c@{}}
\toprule
& Representative monomial & $n$ \\
\midrule
1 & $\eta_{\lambda\mu}\eta_{\nu\rho}\eta_{\sigma\tau}$ & 4 \\
2 & $\eta_{\lambda\mu}\eta_{\nu\sigma}\eta_{\rho\tau}$ & 4 \\
3 & $\eta_{\lambda\rho}\eta_{\mu\nu}\eta_{\sigma\tau}$ & 1 \\
4 & $\eta_{\lambda\rho}\eta_{\mu\sigma}\eta_{\nu\tau}$ & 2 \\
5 & $\eta_{\lambda\sigma}\eta_{\mu\rho}\eta_{\nu\tau}$ & 4 \\
\bottomrule
\end{tabular}
\caption{Basis of $K^{GG}_{\lambda\mu\nu\rho\sigma\tau}$: 5 elements, the orbits of the 15
monomials under the symmetry group of order 8.}
\label{tab:basis-GG}
\end{table}

\begin{table}[htb]
\centering
\setlength{\aboverulesep}{0pt}\setlength{\belowrulesep}{0pt}\renewcommand{\arraystretch}{1.15}
\begin{tabular}{@{}cl c@{\hspace{5pt}}|@{\hspace{5pt}}cl c@{}}
\toprule
& Representative monomial & $n$ & & Representative monomial & $n$ \\
\midrule
1 & $\eta_{\mu\nu}\eta_{\lambda\alpha}\eta_{\beta\rho}\eta_{\sigma\tau}$ & 4 & 9 & $\eta_{\mu\lambda}\eta_{\nu\rho}\eta_{\alpha\sigma}\eta_{\beta\tau}$ & 4 \\
2 & $\eta_{\mu\nu}\eta_{\lambda\alpha}\eta_{\beta\sigma}\eta_{\rho\tau}$ & 4 & 10 & $\eta_{\mu\lambda}\eta_{\nu\sigma}\eta_{\alpha\beta}\eta_{\rho\tau}$ & 8 \\
3 & $\eta_{\mu\nu}\eta_{\lambda\rho}\eta_{\alpha\beta}\eta_{\sigma\tau}$ & 1 & 11 & $\eta_{\mu\lambda}\eta_{\nu\sigma}\eta_{\alpha\rho}\eta_{\beta\tau}$ & 16 \\
4 & $\eta_{\mu\nu}\eta_{\lambda\rho}\eta_{\alpha\sigma}\eta_{\beta\tau}$ & 2 & 12 & $\eta_{\mu\alpha}\eta_{\nu\beta}\eta_{\lambda\rho}\eta_{\sigma\tau}$ & 4 \\
5 & $\eta_{\mu\nu}\eta_{\lambda\sigma}\eta_{\alpha\rho}\eta_{\beta\tau}$ & 4 & 13 & $\eta_{\mu\alpha}\eta_{\nu\beta}\eta_{\lambda\sigma}\eta_{\rho\tau}$ & 8 \\
6 & $\eta_{\mu\lambda}\eta_{\nu\alpha}\eta_{\beta\rho}\eta_{\sigma\tau}$ & 8 & 14 & $\eta_{\mu\alpha}\eta_{\nu\sigma}\eta_{\lambda\beta}\eta_{\rho\tau}$ & 8 \\
7 & $\eta_{\mu\lambda}\eta_{\nu\alpha}\eta_{\beta\sigma}\eta_{\rho\tau}$ & 16 & 15 & $\eta_{\mu\alpha}\eta_{\nu\sigma}\eta_{\lambda\rho}\eta_{\beta\tau}$ & 8 \\
8 & $\eta_{\mu\lambda}\eta_{\nu\rho}\eta_{\alpha\beta}\eta_{\sigma\tau}$ & 2 & 16 & $\eta_{\mu\alpha}\eta_{\nu\sigma}\eta_{\lambda\tau}\eta_{\beta\rho}$ & 8 \\
\bottomrule
\end{tabular}
\caption{Basis of $K^{\phi GG}_{\mu\nu\lambda\alpha\beta\rho\sigma\tau}$: 16 elements, the
orbits of the 105 monomials (one for each way of pairing the eight indices into four
$\eta$'s) under the symmetry group of order 16.}
\label{tab:basis-phiGG}
\end{table}

\section{The six identities}
\label{app:identities}

In the notation of Section~\ref{sec:variables}, each identity of Table~\ref{tab:identities}
states that an expression linear in each field and in the gauge parameter vanishes identically.
Writing $P$ for the total momentum ($P=p_1+p_2$ or $p_1+p_2+p_3$), the six forms are
\begin{align}
\W^{\phi}:&\quad 2K^{\phi\phi}\big(\phi_{-p},D_{\phi^0}(p)\,\omega_p\big)
+K^{\phi G}\big(\phi_{-p},D_{G^0}(p)\,\omega_p\big)=0,\\
\W^{G}:&\quad K^{\phi G}\big(D_{\phi^0}(p)\,\omega_p,G_{-p}\big)
+2K^{GG}\big(G_{-p},D_{G^0}(p)\,\omega_p\big)=0,\\
\W^{31}:&\quad K^{\phi\phi}\big(\phi_{-p_1},D_{\phi^1}(p_1,P)(\phi_{-p_2},\omega_P)\big)
+(1\leftrightarrow2)=0,\\
\begin{split}
\W^{32}:&\quad 2K^{\phi GG}\big(\phi_{-p_2},D_{G^0}(P)\,\omega_P,G_{-p_1}\big)
+K^{\phi G}\big(D_{\phi^1}(p_1,P)(\phi_{-p_2},\omega_P),G_{-p_1}\big)\\
&\quad+K^{\phi G}\big(\phi_{-p_2},D_{G^1}(p_2,P)(G_{-p_1},\omega_P)\big)=0,
\end{split}\\
\begin{split}
\W^{33}:&\quad K^{\phi GG}\big(D_{\phi^0}(P)\,\omega_P,G_{-p_1},G_{-p_2}\big)
+K^{GG}\big(G_{-p_1},D_{G^1}(p_1,P)(G_{-p_2},\omega_P)\big)+(1\leftrightarrow2)=0,
\end{split}\\
\begin{split}
\W^{34}:&\quad K^{\phi GG}\big(\phi_{-p_3},G_{-p_1},D_{G^1}(p_1+p_3,P)(G_{-p_2},\omega_P)\big)
+(1\leftrightarrow2)\\
&\quad+K^{\phi GG}\big(D_{\phi^1}(p_1+p_2,P)(\phi_{-p_3},\omega_P),G_{-p_1},G_{-p_2}\big)=0,
\end{split}
\end{align}
where $(1\leftrightarrow2)$ adds the immediately preceding term with the two field
arguments of the same type exchanged, that is, with $p_1\leftrightarrow p_2$ and, in the
component form below, with their index groups exchanged as well; in $\W^{33}$ it applies
only to the $K^{GG}$ term. The factors of two count the two equal contributions of a kernel
that is quadratic in the field being varied. Each term is one kernel of
Eq.~\eqref{eq:ansatz} with one argument replaced by the appropriate piece of the
transformation \eqref{eq:dmom}, the momenta being fixed by momentum conservation. Since the
fields are arbitrary within their symmetry classes, the coefficient tensor of
$\phi_{-p_1}\cdots\omega_P$ in each form, projected onto the index symmetries of the field
indices, vanishes. The scalar forms above only define the identities and are not used as
such: we work throughout with these coefficient tensors, whose free indices are kept so that
each identity can be projected onto the bases described below. With all indices
displayed, and in terms
of the tensors of Appendix~\ref{app:momentum}, these coefficient tensors are
\begin{align}
\W^{\phi}_{\rho\sigma\alpha}(p)&\equiv
2D_{\phi^0}{}^{\mu\nu}{}_{\alpha}(p)K^{\phi\phi}_{\mu\nu\rho\sigma}(p)
+D_{G^0}{}^{\lambda}{}_{\mu\nu\alpha}(p)K^{\phi G}_{\rho\sigma\lambda}{}^{\mu\nu}(p)=0,\\
\W^{G}_{\rho\sigma\tau\alpha}(p)&\equiv
D_{\phi^0}{}^{\mu\nu}{}_{\alpha}(p)K^{\phi G}_{\mu\nu\rho\sigma\tau}(-p)
+2D_{G^0}{}^{\lambda}{}_{\mu\nu\alpha}(p)K^{GG}_{\lambda}{}^{\mu\nu}{}_{\rho\sigma\tau}=0,\\
\begin{split}
\W^{31}_{\mu_1\nu_1\mu_2\nu_2\rho}(p_1,p_2)&\equiv
K^{\phi\phi}_{\mu_1\nu_1\alpha\beta}(p_1)D_{\phi^1}{}^{\alpha\beta}{}_{\mu_2\nu_2\rho}(p_1,p_1+p_2)\\
&\quad+K^{\phi\phi}_{\mu_2\nu_2\alpha\beta}(p_2)D_{\phi^1}{}^{\alpha\beta}{}_{\mu_1\nu_1\rho}(p_2,p_1+p_2)=0,
\end{split}\\
\begin{split}
\W^{32}_{\lambda_1\mu_1\nu_1\mu_2\nu_2\rho}(p_1,p_2)&\equiv
2K^{\phi GG}_{\mu_2\nu_2\lambda\mu\nu\lambda_1\mu_1\nu_1}D_{G^0}{}^{\lambda\mu\nu}{}_{\rho}(p_1+p_2)\\
&\quad+K^{\phi G}_{\mu\nu\lambda_1\mu_1\nu_1}(-p_1)D_{\phi^1}{}^{\mu\nu}{}_{\mu_2\nu_2\rho}(p_1,p_1+p_2)\\
&\quad+K^{\phi G}_{\mu_2\nu_2\lambda}{}^{\mu\nu}(p_2)D_{G^1}{}^{\lambda}{}_{\mu\nu\lambda_1\mu_1\nu_1\rho}(p_2,p_1+p_2)=0,
\end{split}\\
\begin{split}
\W^{33}_{\lambda_1\mu_1\nu_1\lambda_2\mu_2\nu_2\rho}(p_1,p_2)&\equiv
K^{\phi GG}_{\mu\nu\lambda_1\mu_1\nu_1\lambda_2\mu_2\nu_2}D_{\phi^0}{}^{\mu\nu}{}_{\rho}(p_1+p_2)\\
&\quad+K^{GG}_{\lambda\mu\nu\lambda_1\mu_1\nu_1}D_{G^1}{}^{\lambda\mu\nu}{}_{\lambda_2\mu_2\nu_2\rho}(p_1,p_1+p_2)\\
&\quad+K^{GG}_{\lambda\mu\nu\lambda_2\mu_2\nu_2}D_{G^1}{}^{\lambda\mu\nu}{}_{\lambda_1\mu_1\nu_1\rho}(p_2,p_1+p_2)=0,
\end{split}\\
\begin{split}
\W^{34}_{\lambda_1\mu_1\nu_1\lambda_2\mu_2\nu_2\mu_3\nu_3\rho}(p_1,p_2,p_3)&\equiv\\
&\quad\phantom{+}K^{\phi GG}_{\mu_3\nu_3\lambda\mu\nu\lambda_1\mu_1\nu_1}D_{G^1}{}^{\lambda\mu\nu}{}_{\lambda_2\mu_2\nu_2\rho}(p_1+p_3,P)\\
&\quad+K^{\phi GG}_{\mu_3\nu_3\lambda\mu\nu\lambda_2\mu_2\nu_2}D_{G^1}{}^{\lambda\mu\nu}{}_{\lambda_1\mu_1\nu_1\rho}(p_2+p_3,P)\\
&\quad+K^{\phi GG}_{\mu\nu\lambda_1\mu_1\nu_1\lambda_2\mu_2\nu_2}D_{\phi^1}{}^{\mu\nu}{}_{\mu_3\nu_3\rho}(p_1+p_2,P)=0,
\end{split}
\end{align}
The identities above retain their generic momentum dependence. For the projection bases
of the cubic identities the independent momenta are identified, $p_1=p_2=p$, and also
$p_3=p$ in $\W^{34}$, as explained in Section~\ref{sec:identities}. The projection bases of
Table~\ref{tab:identities} are generated by the procedure of Appendix~\ref{app:bases}, from
the monomials with the index structure and degree in $p$ of each identity, averaged over its
symmetry group: the symmetries of the kernel indices involved
and, for $\W^{31}$, $\W^{33}$ and $\W^{34}$, the exchange of the two identical fields.
Table~\ref{tab:proj} lists the three smallest ones as an illustration; the remaining
dimensions, 23, 20 and 95, make an explicit listing unhelpful, and those bases are generated
by the code of Appendix~\ref{app:implementation}. Their Gram determinants are, up to
numerical factors, $(p^2)^9(d-1)^2$, $(p^2)^{12}(d-2)(d-1)^5(d+2)(d+4)$ and
$(p^2)^{21}(d-2)^2(d-1)^6(d+1)(d+4)(d+6)$ for the three bases of Table~\ref{tab:proj}, and
for $p^2\neq0$ those of the bases of $\W^{32}$ and $\W^{33}$ vanish, among positive
integer dimensions, only at $d=1,2,3$; all five are therefore nondegenerate for $d\geq4$ and
$p^2\neq0$. The basis of $\W^{34}$ becomes degenerate at $d=4$. This does not obstruct the
determination, since no inversion of that Gram matrix is needed: the other identities fix
the kernels, $\W^{34}$ is then found to be satisfied identically by substitution, and its
vanishing also follows from the identification of the solution with the invariant
Goldberg--Palatini action.

\begin{table}[htb]
\centering\footnotesize
\setlength{\aboverulesep}{0pt}\setlength{\belowrulesep}{0pt}\renewcommand{\arraystretch}{1.15}
\begin{tabular}{@{}cl c@{\hspace{5pt}}|@{\hspace{5pt}}cl c@{}}
\toprule
& Representative monomial & $n$ & & Representative monomial & $n$ \\
\midrule
\multicolumn{3}{@{}l@{\hspace{5pt}}|@{\hspace{5pt}}}{$\W^{\phi}_{\rho\sigma\alpha}(p)$} & \multicolumn{3}{l@{}}{$\W^{31}_{\mu_1\nu_1\mu_2\nu_2\rho}(p,p)$} \\
1 & $p_{\rho}p_{\sigma}p_{\alpha}$ & 1 & 1 & $p_{\mu_1}p_{\nu_1}p_{\rho}\eta_{\mu_2\nu_2}$ & 2 \\
2 & $p^2\,p_{\alpha}\eta_{\rho\sigma}$ & 1 & 2 & $p_{\mu_1}p_{\nu_1}p_{\mu_2}\eta_{\nu_2\rho}$ & 4 \\
3 & $p^2\,p_{\rho}\eta_{\sigma\alpha}$ & 2 & 3 & $p_{\mu_1}p_{\mu_2}p_{\rho}\eta_{\nu_1\nu_2}$ & 4 \\
\multicolumn{3}{@{}l@{\hspace{5pt}}|@{\hspace{5pt}}}{$\W^{G}_{\rho\sigma\tau\alpha}(p)$} & 4 & $p^2\,p_{\rho}\eta_{\mu_1\nu_1}\eta_{\mu_2\nu_2}$ & 1 \\
1 & $p_{\rho}p_{\alpha}\eta_{\sigma\tau}$ & 1 & 5 & $p^2\,p_{\rho}\eta_{\mu_1\mu_2}\eta_{\nu_1\nu_2}$ & 2 \\
2 & $p_{\sigma}p_{\tau}\eta_{\rho\alpha}$ & 1 & 6 & $p^2\,p_{\mu_1}\eta_{\nu_1\rho}\eta_{\mu_2\nu_2}$ & 4 \\
3 & $p_{\rho}p_{\sigma}\eta_{\tau\alpha}$ & 2 & 7 & $p^2\,p_{\mu_1}\eta_{\nu_1\mu_2}\eta_{\nu_2\rho}$ & 8 \\
4 & $p_{\sigma}p_{\alpha}\eta_{\rho\tau}$ & 2 & \multicolumn{3}{l@{}}{} \\
5 & $p^2\,\eta_{\rho\alpha}\eta_{\sigma\tau}$ & 1 & \multicolumn{3}{l@{}}{} \\
6 & $p^2\,\eta_{\rho\sigma}\eta_{\tau\alpha}$ & 2 & \multicolumn{3}{l@{}}{} \\
\bottomrule
\end{tabular}
\caption{The three smallest projection bases, in the notation of Appendix~\ref{app:bases}:
3, 6 and 7 elements, the orbits of 4, 9 and 25 monomials under symmetry groups of order 2, 2
and 8. For $\W^{31}$ the two momenta are identified, $p_1=p_2=p$, and the symmetry group
includes the exchange $(\mu_1\nu_1)\leftrightarrow(\mu_2\nu_2)$.}
\label{tab:proj}
\end{table}

\section{Explicit kernels}
\label{app:kernels}

Up to the overall constant, the solution of Table~\ref{tab:flow} is, in terms of the basis
elements $\mathcal{S}[T]$ of Appendix~\ref{app:bases} (the average of the monomial $T$ over
its orbit), and with all indices of the kernels lowered, to be raised with $\eta^{\mu\nu}$
as required by the contractions with the fields,
\begin{align}
K^{\phi\phi}_{\mu\nu\rho\sigma}(p)&=0,\\
K^{\phi G}_{\mu\nu\lambda\rho\sigma}(p)&=-i\,\mathcal{S}\big[p_\lambda\eta_{\mu\rho}\eta_{\nu\sigma}\big]
=-\tfrac{i}{2}\,p_\lambda\big(\eta_{\mu\rho}\eta_{\nu\sigma}+\eta_{\mu\sigma}\eta_{\nu\rho}\big),\\
\begin{split}
K^{GG}_{\lambda\mu\nu\rho\sigma\tau}&=
-\mathcal{S}\big[\eta_{\lambda\sigma}\eta_{\mu\rho}\eta_{\nu\tau}\big]
+\frac{1}{d-1}\,\mathcal{S}\big[\eta_{\lambda\mu}\eta_{\nu\sigma}\eta_{\rho\tau}\big]\\
&=\tfrac14\Big[-\eta_{\lambda(\sigma}\eta_{\tau)(\mu}\eta_{\nu)\rho}
+\frac{1}{d-1}\,\eta_{\lambda(\mu}\eta_{\nu)(\sigma}\eta_{\tau)\rho}\Big],
\end{split}\\
K^{\phi GG}_{\mu\nu\lambda\alpha\beta\rho\sigma\tau}&=
-\mathcal{S}\big[\eta_{\mu\alpha}\eta_{\nu\sigma}\eta_{\lambda\tau}\eta_{\beta\rho}\big]
+\frac{1}{d-1}\,\mathcal{S}\big[\eta_{\mu\alpha}\eta_{\nu\sigma}\eta_{\lambda\beta}\eta_{\rho\tau}\big],
\end{align}
that is, element~4 of Table~\ref{tab:basis-phiG}, elements 5 and 2 of
Table~\ref{tab:basis-GG}, and elements 16 and 14 of Table~\ref{tab:basis-phiGG}; the latter
two basis elements contain eight terms each. In the $GG$ and $\phi GG$ kernels the first
structure contracts the connection index of one $G$ with a lower index of the other, and the
second is the product of the two traces $g_\mu\equiv G^{\sigma}{}_{\mu\sigma}$.
Contracted with the fields, in the notation of Section~\ref{sec:variables}, these give
\begin{align}
K^{\phi G}(\phi_{-p},G_p)&=-ip_\lambda\,\p^{\mu\nu}(-p)\G^{\lambda}{}_{\mu\nu}(p),\\
K^{GG}(G_{-p},G_p)&=
\eta^{\mu\nu}\Big(-\G^{\lambda}{}_{\mu\sigma}(-p)\G^{\sigma}{}_{\nu\lambda}(p)+\frac{1}{d-1}\G^{\sigma}{}_{\mu\sigma}(-p)\G^{\rho}{}_{\nu\rho}(p)\Big),\\
K^{\phi GG}(\phi_{p+q},G_{-p},G_{-q})&=
\p^{\mu\nu}(p+q)\Big(-\G^{\lambda}{}_{\mu\sigma}(-p)\G^{\sigma}{}_{\nu\lambda}(-q)+\frac{1}{d-1}\G^{\sigma}{}_{\mu\sigma}(-p)\G^{\rho}{}_{\nu\rho}(-q)\Big),
\end{align}
whose inverse Fourier transform yields Eq.~\eqref{eq:result}.

\section{Symbolic implementation and reproducibility}
\label{app:implementation}

In the first-order bootstrap implementation the coupling is set to $\kappa=1$. Since
$\kappa$ enters the construction only as the overall factor of $S_3$ in Eq.~\eqref{eq:ansatz}
and of $\delta^{(1)}$ in Eq.~\eqref{eq:dmom}, this affects neither the kernels of
Appendix~\ref{app:kernels} nor the identities of Appendix~\ref{app:identities}; it is restored
in the results by dimensional analysis, as in Section~\ref{sec:closure}. The second-order
comparison keeps $\kappa$ symbolic throughout.

\paragraph{Files.} The calculations are organized in the following Mathematica~\cite{Mathematica}
notebooks and data files, kept together in one directory.
\begin{enumerate}
\item \texttt{Goldberg\_Palatini\_Bootstrap\_en.nb} (FeynCalc~\cite{FeynCalc}): the first-order
determination. It generates the tensor bases of the four kernels and of the six identities
(Appendices~\ref{app:bases} and \ref{app:identities}) and the Gram determinants of the kernel
bases and of the five smallest projection bases (the rank test of the 95-element basis of
$\W^{34}$ at $d=4$ is an optional diagnostic, not part of the main flow), builds the
nonlinear identities with the field momenta identified, solves the linear system in the sequence of
Table~\ref{tab:flow}, checks the resulting kernels against Eqs.~(3.5)--(3.6) of
Ref.~\cite{BrandtMcKeon2016}, verifies the complete cubic kernel against $\tfrac12M(\phi)$
(cell tagged \texttt{FullCubicKernelCheck}), and exports the normalized kernels.
\item \texttt{Goldberg\_Palatini\_Bootstrap\_kernels.wl}: the exported kernels $K^{GG}$ and
$K^{\phi GG}$ in a package-independent representation by index positions, with the conventions
(dimension symbol, symmetries, normalization $c_{GG}=1/(d-1)$, Fourier sign); it is data,
not an executable script.
\item \texttt{Goldberg\_Palatini\_Cubic\_CLM.nb} (xAct/xTensor, xTras and xPerm~\cite{xAct,xPerm,xTras}):
the second-order comparison of Section~\ref{sec:checks}. It imports the data file without
FeynCalc, converts it to xAct tensors, derives $M_0^{-1}$ by solving its defining identity in
a five-element basis of metric structures and checks both inverse relations
(\texttt{inverseChecks}), substitutes $G_0$ back into $M_0G_0=\partial\phi$ as a test of index
placement and normalization (\texttt{g0Check}), and evaluates the residuals of
Eqs.~\eqref{eq:clm}
(\texttt{clmQuadraticCheck}, \texttt{quadraticBoundaryCheck}, \texttt{clmCubicResidual});
\texttt{allChecksPassed} only aggregates these residuals. A final section compares, in terms of
the full density $h^{\mu\nu}$, the action obtained by eliminating $G$ with Eq.~(A14) of
Ref.~\cite{CLM1973}. The notebook uses the
metric-compatible derivative of the flat background, \texttt{CD}, which in Cartesian
coordinates is the partial derivative of the text; the dimension is kept symbolic, and the
pole of $M_0^{-1}$ at $d=2$ is separate from the range $d\geq4$ of the bases.
\end{enumerate}
A \texttt{README} gives the evaluation order (the first notebook and its export cell first;
the comparison notebook in a fresh kernel), the required packages and versions, the expected
result of every check, the conventions of the comparison, and the correspondence between the
notebooks and the equations and tables of this paper.

The calculations were carried out with Mathematica~15.0.0 (macOS) and repeated with
Mathematica~13.3.1 (macOS) and~14.2.0 (Linux), with identical results, using FeynCalc~10.1.0,
xTensor~1.2.0, xPerm~1.2.3 and xTras~1.4.2; the notebooks require Mathematica and these
packages, and other versions have not been systematically tested. The notebooks, data file, README and the source
of this paper are archived on Zenodo at \url{https://doi.org/10.5281/zenodo.22757769}.

\bibliographystyle{apsrev4-2}
\bibliography{bootstrap_refs_v3}

\begin{thebibliography}{20}%
\makeatletter
\providecommand \@ifxundefined [1]{%
 \@ifx{#1\undefined}
}%
\providecommand \@ifnum [1]{%
 \ifnum #1\expandafter \@firstoftwo
 \else \expandafter \@secondoftwo
 \fi
}%
\providecommand \@ifx [1]{%
 \ifx #1\expandafter \@firstoftwo
 \else \expandafter \@secondoftwo
 \fi
}%
\providecommand \natexlab [1]{#1}%
\providecommand \enquote  [1]{``#1''}%
\providecommand \bibnamefont  [1]{#1}%
\providecommand \bibfnamefont [1]{#1}%
\providecommand \citenamefont [1]{#1}%
\providecommand \href@noop [0]{\@secondoftwo}%
\providecommand \href [0]{\begingroup \@sanitize@url \@href}%
\providecommand \@href[1]{\@@startlink{#1}\@@href}%
\providecommand \@@href[1]{\endgroup#1\@@endlink}%
\providecommand \@sanitize@url [0]{\catcode `\\12\catcode `\$12\catcode
  `\&12\catcode `\#12\catcode `\^12\catcode `\_12\catcode `\%12\relax}%
\providecommand \@@startlink[1]{}%
\providecommand \@@endlink[0]{}%
\providecommand \url  [0]{\begingroup\@sanitize@url \@url }%
\providecommand \@url [1]{\endgroup\@href {#1}{\urlprefix }}%
\providecommand \urlprefix  [0]{URL }%
\providecommand \Eprint [0]{\href }%
\providecommand \doibase [0]{https://doi.org/}%
\providecommand \selectlanguage [0]{\@gobble}%
\providecommand \bibinfo  [0]{\@secondoftwo}%
\providecommand \bibfield  [0]{\@secondoftwo}%
\providecommand \translation [1]{[#1]}%
\providecommand \BibitemOpen [0]{}%
\providecommand \bibitemStop [0]{}%
\providecommand \bibitemNoStop [0]{.\EOS\space}%
\providecommand \EOS [0]{\spacefactor3000\relax}%
\providecommand \BibitemShut  [1]{\csname bibitem#1\endcsname}%
\let\auto@bib@innerbib\@empty
\bibitem [{\citenamefont {Feynman}\ \emph {et~al.}(1995)\citenamefont
  {Feynman}, \citenamefont {Morinigo},\ and\ \citenamefont
  {Wagner}}]{Feynman1995}%
  \BibitemOpen
  \bibfield  {author} {\bibinfo {author} {\bibfnamefont {R.~P.}\ \bibnamefont
  {Feynman}}, \bibinfo {author} {\bibfnamefont {F.~B.}\ \bibnamefont
  {Morinigo}},\ and\ \bibinfo {author} {\bibfnamefont {W.~G.}\ \bibnamefont
  {Wagner}},\ }\href@noop {} {\emph {\bibinfo {title} {Feynman Lectures on
  Gravitation}}},\ edited by\ \bibinfo {editor} {\bibfnamefont
  {B.}~\bibnamefont {Hatfield}}\ (\bibinfo  {publisher} {Addison-Wesley},\
  \bibinfo {address} {Reading, MA},\ \bibinfo {year} {1995})\BibitemShut
  {NoStop}%
\bibitem [{\citenamefont {Deser}(1970)}]{Deser1970}%
  \BibitemOpen
  \bibfield  {author} {\bibinfo {author} {\bibfnamefont {S.}~\bibnamefont
  {Deser}},\ }\href {https://doi.org/10.1007/BF00759198} {\bibfield  {journal}
  {\bibinfo  {journal} {Gen. Rel. Grav.}\ }\textbf {\bibinfo {volume} {1}},\
  \bibinfo {pages} {9} (\bibinfo {year} {1970})},\ \Eprint
  {https://arxiv.org/abs/gr-qc/0411023} {arXiv:gr-qc/0411023} \BibitemShut
  {NoStop}%
\bibitem [{\citenamefont {Palatini}(1919)}]{Palatini1919}%
  \BibitemOpen
  \bibfield  {author} {\bibinfo {author} {\bibfnamefont {A.}~\bibnamefont
  {Palatini}},\ }\href {https://doi.org/10.1007/BF03014670} {\bibfield
  {journal} {\bibinfo  {journal} {Rend. Circ. Mat. Palermo}\ }\textbf {\bibinfo
  {volume} {43}},\ \bibinfo {pages} {203} (\bibinfo {year} {1919})}\BibitemShut
  {NoStop}%
\bibitem [{\citenamefont {Ferraris}\ \emph {et~al.}(1982)\citenamefont
  {Ferraris}, \citenamefont {Francaviglia},\ and\ \citenamefont
  {Reina}}]{Ferraris1982}%
  \BibitemOpen
  \bibfield  {author} {\bibinfo {author} {\bibfnamefont {M.}~\bibnamefont
  {Ferraris}}, \bibinfo {author} {\bibfnamefont {M.}~\bibnamefont
  {Francaviglia}},\ and\ \bibinfo {author} {\bibfnamefont {C.}~\bibnamefont
  {Reina}},\ }\href {https://doi.org/10.1007/BF00756060} {\bibfield  {journal}
  {\bibinfo  {journal} {Gen. Rel. Grav.}\ }\textbf {\bibinfo {volume} {14}},\
  \bibinfo {pages} {243} (\bibinfo {year} {1982})}\BibitemShut {NoStop}%
\bibitem [{\citenamefont {Padmanabhan}(2008)}]{Padmanabhan2008}%
  \BibitemOpen
  \bibfield  {author} {\bibinfo {author} {\bibfnamefont {T.}~\bibnamefont
  {Padmanabhan}},\ }\href {https://doi.org/10.1142/S0218271808012085}
  {\bibfield  {journal} {\bibinfo  {journal} {Int. J. Mod. Phys. D}\ }\textbf
  {\bibinfo {volume} {17}},\ \bibinfo {pages} {367} (\bibinfo {year} {2008})},\
  \Eprint {https://arxiv.org/abs/gr-qc/0409089} {arXiv:gr-qc/0409089}
  \BibitemShut {NoStop}%
\bibitem [{\citenamefont {Deser}(2010)}]{Deser2010}%
  \BibitemOpen
  \bibfield  {author} {\bibinfo {author} {\bibfnamefont {S.}~\bibnamefont
  {Deser}},\ }\href {https://doi.org/10.1007/s10714-009-0912-9} {\bibfield
  {journal} {\bibinfo  {journal} {Gen. Rel. Grav.}\ }\textbf {\bibinfo {volume}
  {42}},\ \bibinfo {pages} {641} (\bibinfo {year} {2010})},\ \Eprint
  {https://arxiv.org/abs/0910.2975} {arXiv:0910.2975} \BibitemShut {NoStop}%
\bibitem [{\citenamefont {Brandt}\ and\ \citenamefont
  {Elias-Filho}(2006)}]{BrandtElias2006}%
  \BibitemOpen
  \bibfield  {author} {\bibinfo {author} {\bibfnamefont {F.~T.}\ \bibnamefont
  {Brandt}}\ and\ \bibinfo {author} {\bibfnamefont {M.~R.}\ \bibnamefont
  {Elias-Filho}},\ }\href {https://doi.org/10.1103/PhysRevD.74.067704}
  {\bibfield  {journal} {\bibinfo  {journal} {Phys. Rev. D}\ }\textbf {\bibinfo
  {volume} {74}},\ \bibinfo {pages} {067704} (\bibinfo {year} {2006})},\
  \Eprint {https://arxiv.org/abs/hep-th/0609106} {arXiv:hep-th/0609106}
  \BibitemShut {NoStop}%
\bibitem [{\citenamefont {Brandt}\ and\ \citenamefont
  {Frenkel}(1996)}]{BrandtFrenkel1996}%
  \BibitemOpen
  \bibfield  {author} {\bibinfo {author} {\bibfnamefont {F.~T.}\ \bibnamefont
  {Brandt}}\ and\ \bibinfo {author} {\bibfnamefont {J.}~\bibnamefont
  {Frenkel}},\ }\href {https://doi.org/10.1103/PhysRevD.53.911} {\bibfield
  {journal} {\bibinfo  {journal} {Phys. Rev. D}\ }\textbf {\bibinfo {volume}
  {53}},\ \bibinfo {pages} {911} (\bibinfo {year} {1996})},\ \Eprint
  {https://arxiv.org/abs/hep-th/9505114} {arXiv:hep-th/9505114} \BibitemShut
  {NoStop}%
\bibitem [{\citenamefont {Goldberg}(1958)}]{Goldberg1958}%
  \BibitemOpen
  \bibfield  {author} {\bibinfo {author} {\bibfnamefont {J.~N.}\ \bibnamefont
  {Goldberg}},\ }\href {https://doi.org/10.1103/PhysRev.111.315} {\bibfield
  {journal} {\bibinfo  {journal} {Phys. Rev.}\ }\textbf {\bibinfo {volume}
  {111}},\ \bibinfo {pages} {315} (\bibinfo {year} {1958})}\BibitemShut
  {NoStop}%
\bibitem [{\citenamefont {Brandt}\ and\ \citenamefont
  {McKeon}(2016)}]{BrandtMcKeon2016}%
  \BibitemOpen
  \bibfield  {author} {\bibinfo {author} {\bibfnamefont {F.~T.}\ \bibnamefont
  {Brandt}}\ and\ \bibinfo {author} {\bibfnamefont {D.~G.~C.}\ \bibnamefont
  {McKeon}},\ }\href {https://doi.org/10.1103/PhysRevD.93.105037} {\bibfield
  {journal} {\bibinfo  {journal} {Phys. Rev. D}\ }\textbf {\bibinfo {volume}
  {93}},\ \bibinfo {pages} {105037} (\bibinfo {year} {2016})},\ \Eprint
  {https://arxiv.org/abs/1601.04944} {arXiv:1601.04944} \BibitemShut {NoStop}%
\bibitem [{\citenamefont {Delhom}\ \emph {et~al.}(2023)\citenamefont {Delhom},
  \citenamefont {Garc{\'\i}a-Moreno}, \citenamefont {Hohmann}, \citenamefont
  {Jim{\'e}nez-Cano},\ and\ \citenamefont {Koivisto}}]{Delhom2023}%
  \BibitemOpen
  \bibfield  {author} {\bibinfo {author} {\bibfnamefont {A.}~\bibnamefont
  {Delhom}}, \bibinfo {author} {\bibfnamefont {G.}~\bibnamefont
  {Garc{\'\i}a-Moreno}}, \bibinfo {author} {\bibfnamefont {M.}~\bibnamefont
  {Hohmann}}, \bibinfo {author} {\bibfnamefont {A.}~\bibnamefont
  {Jim{\'e}nez-Cano}},\ and\ \bibinfo {author} {\bibfnamefont {T.~S.}\
  \bibnamefont {Koivisto}},\ }\href
  {https://doi.org/10.1088/1475-7516/2023/12/006} {\bibfield  {journal}
  {\bibinfo  {journal} {JCAP}\ }\textbf {\bibinfo {volume} {12}}\bibfield
  {number} {\bibinfo  {number} { (12)},\ \bibinfo {pages} {006}},\ }\Eprint
  {https://arxiv.org/abs/2211.13056} {arXiv:2211.13056} \BibitemShut {NoStop}%
\bibitem [{\citenamefont {Wald}(1986)}]{Wald1986}%
  \BibitemOpen
  \bibfield  {author} {\bibinfo {author} {\bibfnamefont {R.~M.}\ \bibnamefont
  {Wald}},\ }\href {https://doi.org/10.1103/PhysRevD.33.3613} {\bibfield
  {journal} {\bibinfo  {journal} {Phys. Rev. D}\ }\textbf {\bibinfo {volume}
  {33}},\ \bibinfo {pages} {3613} (\bibinfo {year} {1986})}\BibitemShut
  {NoStop}%
\bibitem [{\citenamefont {Brandt}\ and\ \citenamefont
  {McKeon}(2015)}]{BrandtMcKeon2015}%
  \BibitemOpen
  \bibfield  {author} {\bibinfo {author} {\bibfnamefont {F.~T.}\ \bibnamefont
  {Brandt}}\ and\ \bibinfo {author} {\bibfnamefont {D.~G.~C.}\ \bibnamefont
  {McKeon}},\ }\href {https://doi.org/10.1103/PhysRevD.91.105006} {\bibfield
  {journal} {\bibinfo  {journal} {Phys. Rev. D}\ }\textbf {\bibinfo {volume}
  {91}},\ \bibinfo {pages} {105006} (\bibinfo {year} {2015})},\ \Eprint
  {https://arxiv.org/abs/1503.02598} {arXiv:1503.02598} \BibitemShut {NoStop}%
\bibitem [{\citenamefont {Brandt}\ and\ \citenamefont
  {Frenkel}(1993)}]{BrandtFrenkel1993}%
  \BibitemOpen
  \bibfield  {author} {\bibinfo {author} {\bibfnamefont {F.~T.}\ \bibnamefont
  {Brandt}}\ and\ \bibinfo {author} {\bibfnamefont {J.}~\bibnamefont
  {Frenkel}},\ }\href {https://doi.org/10.1103/PhysRevD.47.4688} {\bibfield
  {journal} {\bibinfo  {journal} {Phys. Rev. D}\ }\textbf {\bibinfo {volume}
  {47}},\ \bibinfo {pages} {4688} (\bibinfo {year} {1993})},\ \Eprint
  {https://arxiv.org/abs/hep-ph/9209265} {arXiv:hep-ph/9209265} \BibitemShut
  {NoStop}%
\bibitem [{\citenamefont {Shtabovenko}\ \emph {et~al.}(2025)\citenamefont
  {Shtabovenko}, \citenamefont {Mertig},\ and\ \citenamefont
  {Orellana}}]{FeynCalc}%
  \BibitemOpen
  \bibfield  {author} {\bibinfo {author} {\bibfnamefont {V.}~\bibnamefont
  {Shtabovenko}}, \bibinfo {author} {\bibfnamefont {R.}~\bibnamefont
  {Mertig}},\ and\ \bibinfo {author} {\bibfnamefont {F.}~\bibnamefont
  {Orellana}},\ }\href {https://doi.org/10.1016/j.cpc.2024.109357} {\bibfield
  {journal} {\bibinfo  {journal} {Comput. Phys. Commun.}\ }\textbf {\bibinfo
  {volume} {306}},\ \bibinfo {pages} {109357} (\bibinfo {year}
  {2025})}\BibitemShut {NoStop}%
\bibitem [{\citenamefont {{Wolfram Research, Inc.}}(2026)}]{Mathematica}%
  \BibitemOpen
  \bibfield  {author} {\bibinfo {author} {\bibnamefont {{Wolfram Research,
  Inc.}}},\ }\href@noop {} {\bibinfo {title} {Mathematica, version 15.0}},\
  \bibinfo {howpublished} {Champaign, IL} (\bibinfo {year} {2026})\BibitemShut
  {NoStop}%
\bibitem [{\citenamefont {Capper}\ \emph {et~al.}(1973)\citenamefont {Capper},
  \citenamefont {Leibbrandt},\ and\ \citenamefont
  {Ram{\'o}n~Medrano}}]{CLM1973}%
  \BibitemOpen
  \bibfield  {author} {\bibinfo {author} {\bibfnamefont {D.~M.}\ \bibnamefont
  {Capper}}, \bibinfo {author} {\bibfnamefont {G.}~\bibnamefont {Leibbrandt}},\
  and\ \bibinfo {author} {\bibfnamefont {M.}~\bibnamefont
  {Ram{\'o}n~Medrano}},\ }\href {https://doi.org/10.1103/PhysRevD.8.4320}
  {\bibfield  {journal} {\bibinfo  {journal} {Phys. Rev. D}\ }\textbf {\bibinfo
  {volume} {8}},\ \bibinfo {pages} {4320} (\bibinfo {year} {1973})}\BibitemShut
  {NoStop}%
\bibitem [{\citenamefont {Mart{\'\i}n-Garc{\'\i}a}()}]{xAct}%
  \BibitemOpen
  \bibfield  {author} {\bibinfo {author} {\bibfnamefont {J.~M.}\ \bibnamefont
  {Mart{\'\i}n-Garc{\'\i}a}},\ }\href@noop {} {\bibinfo {title} {{xAct}:
  Efficient tensor computer algebra for the {Wolfram Language}}},\ \bibinfo
  {howpublished} {\url{http://www.xact.es}}\BibitemShut {NoStop}%
\bibitem [{\citenamefont {Mart{\'\i}n-Garc{\'\i}a}(2008)}]{xPerm}%
  \BibitemOpen
  \bibfield  {author} {\bibinfo {author} {\bibfnamefont {J.~M.}\ \bibnamefont
  {Mart{\'\i}n-Garc{\'\i}a}},\ }\href
  {https://doi.org/10.1016/j.cpc.2008.05.009} {\bibfield  {journal} {\bibinfo
  {journal} {Comput. Phys. Commun.}\ }\textbf {\bibinfo {volume} {179}},\
  \bibinfo {pages} {597} (\bibinfo {year} {2008})}\BibitemShut {NoStop}%
\bibitem [{\citenamefont {Nutma}(2014)}]{xTras}%
  \BibitemOpen
  \bibfield  {author} {\bibinfo {author} {\bibfnamefont {T.}~\bibnamefont
  {Nutma}},\ }\href {https://doi.org/10.1016/j.cpc.2014.02.006} {\bibfield
  {journal} {\bibinfo  {journal} {Comput. Phys. Commun.}\ }\textbf {\bibinfo
  {volume} {185}},\ \bibinfo {pages} {1719} (\bibinfo {year}
  {2014})}\BibitemShut {NoStop}%
\end{thebibliography}%

\end{document}